\documentclass[english,11pt,a4paper]{extarticle}
\usepackage[T1]{fontenc}
\usepackage[latin9]{inputenc}
\usepackage{color}
\usepackage{babel}
\usepackage{refstyle}
\usepackage{units}
\usepackage{bm}
\usepackage{pdfpages}
\usepackage{amsmath}
\usepackage{graphicx}
\usepackage[unicode=true,pdfusetitle,
 bookmarks=true,bookmarksnumbered=false,bookmarksopen=false,
 breaklinks=true,pdfborder={0 0 0},pdfborderstyle={},backref=false,colorlinks=true]
 {hyperref}
\hypersetup{
 unicode=true,colorlinks=true,urlcolor=blue,anchorcolor=blue,citecolor=blue,filecolor=blue,linkcolor=blue,menucolor=blue,pagecolor=blue,linktocpage=true,pdfa=true}

\makeatletter

\AtBeginDocument{\providecommand\secref[1]{\ref{sec:#1}}}
\AtBeginDocument{\providecommand\figref[1]{\ref{fig:#1}}}
\AtBeginDocument{\providecommand\subsecref[1]{\ref{subsec:#1}}}
\AtBeginDocument{\providecommand\parref[1]{\ref{par:#1}}}

\RS@ifundefined{subsecref}
  {\newref{subsec}{name = \RSsectxt}}
  {}
\RS@ifundefined{thmref}
  {\def\RSthmtxt{theorem~}\newref{thm}{name = \RSthmtxt}}
  {}
\RS@ifundefined{lemref}
  {\def\RSlemtxt{lemma~}\newref{lem}{name = \RSlemtxt}}
  {}

\usepackage{jinstpub}
\pdfoutput=1

\subheader{\jobname  \\ \today \quad \currenttime}
\title{Polar polarization\\ A new method for polarimetry analysis}

\author[a,1]{D.~Izraeli,\note{Corresponding author.}}
\emailAdd{davidizraeli@post.tau.ac.il}
\author[a,b]{I.~Mardor,}
\author[a]{E.\,O.~Cohen,}
\author[a]{M.~Duer,}
\author[c]{T.\,Y.~Izraeli,}
\author[a,d]{I.~Korover,}
\author[a]{J.~Lichtenstadt,}
\author[a]{and E.~Piasetzky}

\affiliation[a]{School of Physics and Astronomy, Tel Aviv University, Tel Aviv 6997801,
Israel.}
\affiliation[b]{Soreq NRC, Yavne 81800, Israel.}
\affiliation[c]{Guardian Optical Technologies, Derech Hashalom 7, Tel Aviv 6789208, Israel.}
\affiliation[d]{Department of Physics, NRCN, P.O. Box 9001, Beer-Sheva 8419001, Israel.}

\usepackage{braket}
\usepackage{mathrsfs}
\usepackage{slashed}
\usepackage{bm}
\usepackage{datetime}
\allowdisplaybreaks
\usepackage{lineno}
\usepackage{setspace}
\usepackage{graphicx}

\usepackage[dvipsnames]{xcolor}

\@ifundefined{showcaptionsetup}{}{%
 \PassOptionsToPackage{caption=false}{subfig}}
\usepackage{subfig}
\makeatother

\begin{document}
\abstract{We present a novel analysis method for measurements of
polarization transferred in $A(\vec{e},e'\vec{N})$ experiments, which
can be applied to other kinds of polarization measurements as well.
In this method the polarization transfer components are presented
in spherical coordinates using an efficient likelihood numerical maximization
based on an analytic derivation. We also propose a formalism that
accounts for multi-parameter models, and which yields a smooth and
continuous representation of the data (rather than using standard
binning). Applying this method on simulated data generates results
with reduced statistical and systematic uncertainties and enables
revealing physical information that is lost in standard binning of
the data. The obtained results can be compared easily to theoretical
models and other measurements. Furthermore, CPU time is significantly
reduced using this method. }

\maketitle 
\flushbottom

\section{Introduction\label{sec:Introduction}}

Polarization is an important example of an observable extracted from
an ensemble of events, which cannot be obtained from a single one.
Other examples include half-life of unstable particles, and the mass
and width of resonances. Furthermore, polarization transfer may depend
on several variables that do not follow a well-defined function so
that the large data set is not characterized by a single value. This
is in contrast to measurements such as half-life or resonance parameters,
where the entire data set is described by a well-defined function
and the obtained parameters characterize the entire data set. 

For the case of polarization-transfer measurements via $A(\vec{e},e'\vec{N})$,
we propose an analysis method that improves the extraction of the
physical content from the measurement by using a better coordinate
system. We also propose to use a continuous (unbinned) presentation
of the results, and reduce computation time by improving the optimization's
starting point.

Polarization measurements are customarily presented in Cartesian coordinates.
In these coordinates the measured components are mixed and are measured
with less accuracy than the magnitude of the polarization vector and
its direction~\cite{pospischil_thesis}.

Generally, when estimators depend on certain parameters, such as kinematic
variables in the case of polarization-transfer, it is customary to
divide the data into bins~\cite{pospischil_thesis,doria_thesis,degrush_thesis}.
The widths of these bins are usually somewhat arbitrary, set to include
enough data in each bin to yield a reasonable uncertainty, and to
display possible variations of the physical result as a function of
the binned parameter, based on prior estimation of its behavior.

By definition, binning constitutes a compromise regarding the quality
of information that may be extracted from the measurement. It includes
an inherent (and usually wrong) assumption that the values in each
bin are independent of each other, and further, that one single physical
value applies to all data within the bin. Thus it may potentially
conceal real variations within the bin width and result in increased
uncertainties. In addition, binning limits the ability to compare
experimental data with theoretical models or with other experiments
that might be binned differently.

Excessive optimization is necessary in statistical measurements, where
the physical results are the estimators that maximize the likelihood
function of the given observations. Such optimization might be CPU
intensive, especially if the starting point is chosen indiscriminately. 

We propose an analysis method that overcomes the above limitations.
In this method, we extract the polarization data in polar coordinates,
which naturally exhibit the polarization magnitude and angular direction.
We calculate analytically a starting point for the likelihood numerical
maximization, and finally, we analyze all measured data points in
a continuous manner. We assume a polynomial behavior of the estimators
in each bin, and match each bin to its neighbors by demanding continuity
and smoothness at all bin edges.

The paper is laid out as follows: in \secref{The-method} we describe
our new method. We show in \secref{Validity} its validity and effectiveness
using simulated polarization results, and we summarize and conclude
in \secref{Summary-and-Conclusions}.

\section{The method\label{sec:The-method}}

In this chapter we describe our new method for the calculation and
analysis of particle polarization. We start by reviewing the traditional
method of polarization estimation through maximum likelihood, and
show that the common linear approximation can be used as a first step
in an efficient numerical optimization. Next we discuss the choice
of coordinate system and present the estimators for polar polarization
components. Afterwards, we treat the case of varying polarization.
Then we present the tools for comparing measurements with theoretical
calculations. At last, we demonstrate how the entire formalism can
be used in a sophisticated situation.

\subsection{Polarization estimation through maximum likelihood}

In this section we review the traditional way of extracting polarization
using the maximum likelihood method. We then show that the approximated
linear likelihood can be used as an initial step of a exact numerical
optimization. We further show that the derivatives required for the
following steps can be derived analytically, thus reducing the computation
time by approximately two orders of magnitude.

\subsubsection{Exact formalism}

We define the Cartesian coordinates of the polarization transferred
to a proton as:
\begin{equation}
\boldsymbol{P}=\left(P_{x},P_{y},P_{z}\right).
\end{equation}

\begin{figure}
\hspace*{\fill}\subfloat[\label{fig:Spin-precession-of}Spin precession of a proton in a spectrometer.
Adapted from~\cite{doria_thesis}.]{\includegraphics[width=0.9\textwidth]{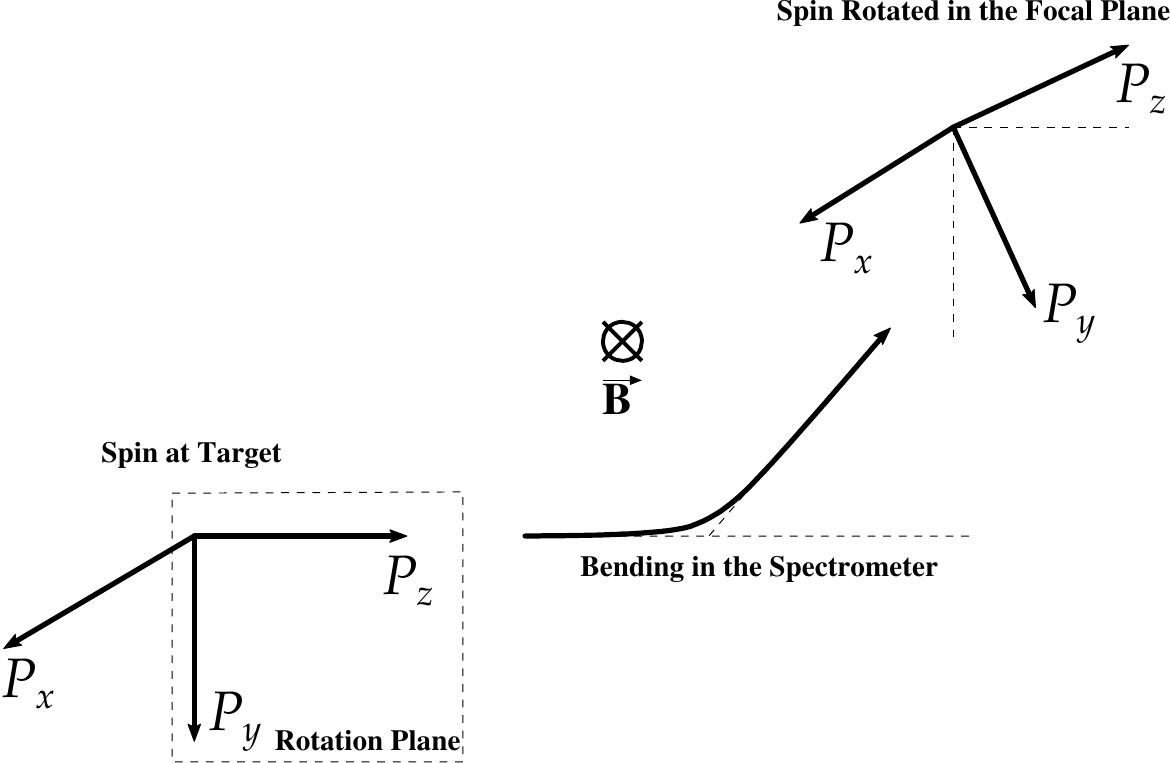}

}\hspace*{\fill}

\hspace*{\fill}\subfloat[\label{fig:phiFpp__Proton-scattering-azimuthal-angle}Proton scattering
azimuthal angle ($\phi_{\mathrm{FPP}}$). Adapted from~\cite{paolone2007polarization}.]{\includegraphics[width=0.9\textwidth]{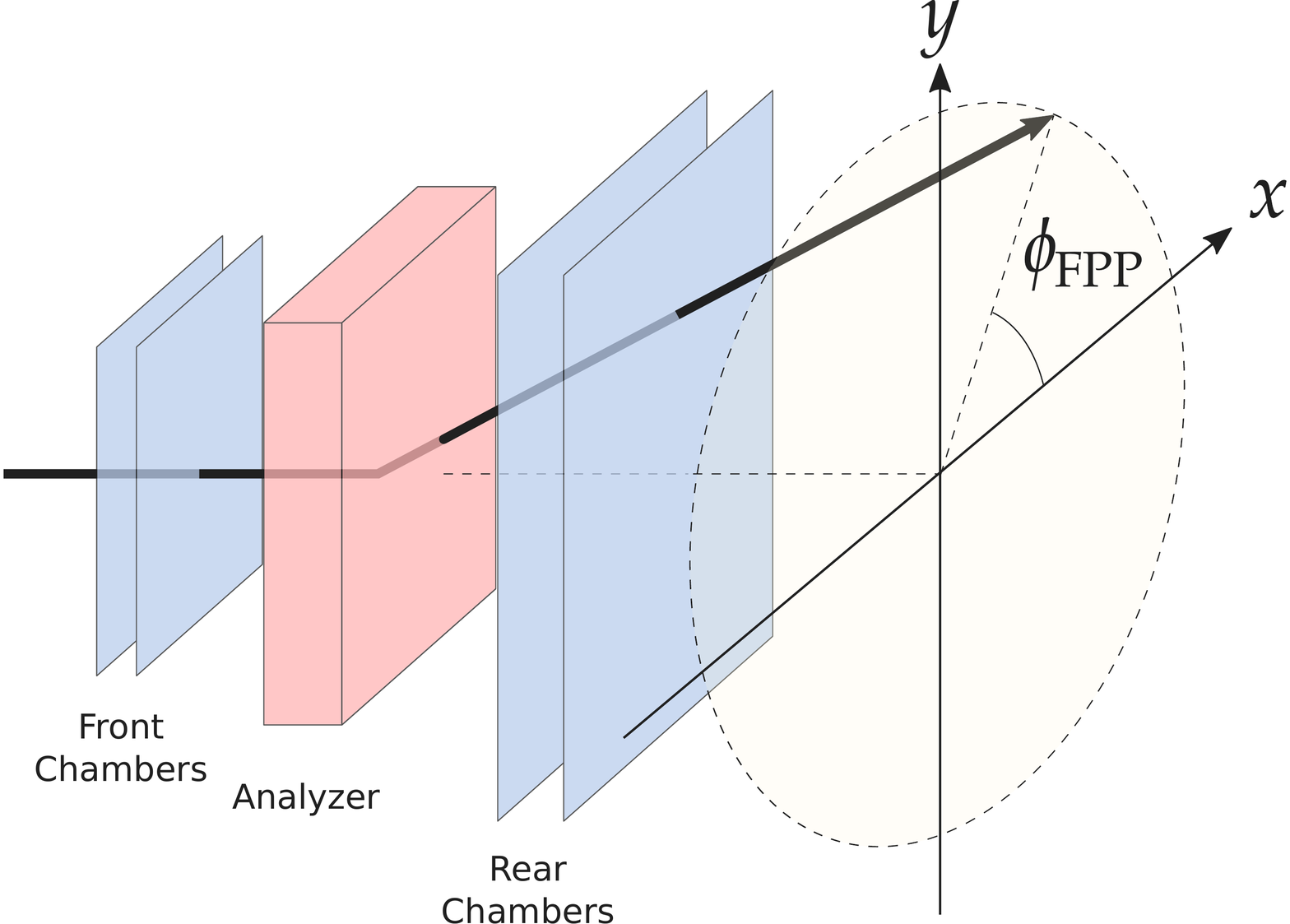}

}\hspace*{\fill}

\caption{Polarization-transfer to a proton measured by a focal plane polarimeter
(FPP). }
\end{figure}

The polarization measured by a focal plane polarimeter (FPP) is determined
from the distribution ($f$) of the proton scattering azimuthal angle
($\phi_{\mathrm{FPP}}$ in \figref{phiFpp__Proton-scattering-azimuthal-angle}),
 as shown in~\cite{BESSET1979}. the distribution is\begin{linenomath}
\begin{equation}
f\left(\phi_{\mathrm{FPP}}\right)=\frac{1+\gamma\left(\phi_{\mathrm{FPP}}\right)}{2\pi},\label{eq:dist}
\end{equation}
\end{linenomath}where $\gamma\left(\phi_{\mathrm{FPP}}\right)$ is
\begin{linenomath}
\begin{align}
\gamma\left(\phi_{\mathrm{FPP}}\right) & \equiv a\;\begin{pmatrix}-\sin\phi_{\mathrm{FPP}}, & \cos\phi_{\mathrm{FPP}}, & 0\end{pmatrix}\mathsf{S}\boldsymbol{P}\nonumber \\
 & =\boldsymbol{P}\cdot\left(a\,\mathsf{S}^{-1}\begin{pmatrix}-\sin\phi_{\mathrm{FPP}}\\
\cos\phi_{\mathrm{FPP}}\\
0
\end{pmatrix}\right)\nonumber \\
 & \equiv\boldsymbol{P}\cdot\bm{{\lambda}}.\label{eq:gamma}
\end{align}
\end{linenomath}Here $\bm{{\lambda}}\equiv a\,\mathsf{S}^{-1}\begin{pmatrix}-\sin\phi_{\mathrm{FPP}}\\
\cos\phi_{\mathrm{FPP}}\\
0
\end{pmatrix}=a\begin{pmatrix}-S_{11}\sin\phi_{\mathrm{FPP}}+S_{21}\cos\phi_{\mathrm{FPP}}\\
-S_{12}\sin\phi_{\mathrm{FPP}}+S_{22}\cos\phi_{\mathrm{FPP}}\\
-S_{13}\sin\phi_{\mathrm{FPP}}+S_{23}\cos\phi_{\mathrm{FPP}}
\end{pmatrix}$, where $S_{ij}=S_{ji}^{-1}$ are the elements of $\mathsf{S}$. We
assume that the sought polarization $\boldsymbol{P}$ is transformed
to the FPP by a known spin precession matrix $\mathsf{S}$ (see \figref{Spin-precession-of}).
In addition we allow for an efficiency factor $0<a<1$ (usually the
product of the analyzing power and the beam polarization).

We calculate the target proton polarization by obtaining its maximum
likelihood estimators (MLE). To this end, we calculate the likelihood
function of the target proton polarization distribution (eq.~(\ref{eq:dist}))
in the standard way: 
\begin{equation}
\mathcal{L}\left(\boldsymbol{P}|\phi_{FPP}\right)=\prod_{k=1}^{n}\left(1+\gamma_{k}\right),
\end{equation}
where we omitted the constant $\left(2\pi\right)^{-n}$, with $n$
the number of events in the sample. As is usually preferred, we continue
the process with the log-likelihood function:
\begin{equation}
L\equiv\ln\mathcal{L}=\sum_{k=1}^{n}\ln\left(1+\gamma_{k}\right).\label{eq:logLike}
\end{equation}

In the following, $\nabla_{\boldsymbol{P}}\equiv\begin{pmatrix}\partial_{P_{x}}\\
\partial_{P_{y}}\\
\partial_{P_{z}}
\end{pmatrix}$ is a column gradient operator, $\nabla_{\boldsymbol{P}}^{T}\equiv\left(\partial_{P_{x}},\partial_{P_{y}},\partial_{P_{z}}\right)$
is a row gradient operator, and the Hessian operator is the $3\times3$
second derivative matrix:\begin{linenomath}
\begin{equation}
\nabla_{\boldsymbol{P}}\otimes\nabla_{\boldsymbol{P}}\equiv\nabla_{\boldsymbol{P}}\nabla_{\boldsymbol{P}}^{T}=\begin{pmatrix}\partial_{P_{x}}^{2} & \partial_{P_{x}}\partial_{P_{y}} & \partial_{P_{x}}\partial_{P_{z}}\\
\partial_{P_{x}}\partial_{P_{y}} & \partial_{P_{y}}^{2} & \partial_{P_{y}}\partial_{P_{z}}\\
\partial_{P_{x}}\partial_{P_{z}} & \partial_{P_{y}}\partial_{P_{z}} & \partial_{P_{z}}^{2}
\end{pmatrix}.
\end{equation}
\end{linenomath}

We extract the MLEs of the polarization by equating the log-likelihood
function\textquoteright s gradient to zero: \begin{linenomath}
\begin{align}
\nabla_{\boldsymbol{P}}L & =\sum_{k=1}^{n}\frac{\nabla_{\boldsymbol{P}}\gamma_{k}}{1+\gamma_{k}}\nonumber \\
 & =\sum_{k=1}^{n}\frac{\bm{{\lambda}}_{k}}{1+\boldsymbol{P}\cdot\bm{{\lambda}}_{k}}\stackrel{!}{=}0,\label{eq:grad}
\end{align}
\end{linenomath}where the second expression, in terms of $\bm{{\lambda}}_{k}$,
is derived using eq.~(\ref{eq:gamma}) and $\nabla_{\boldsymbol{P}}\,\gamma=\bm{{\lambda}}$.
The covariance matrix of the MLEs is given by the inverse of minus
the Hessian of the log-likelihood:\begin{linenomath}
\begin{align}
H & \equiv\nabla_{\boldsymbol{P}}\otimes\nabla_{\boldsymbol{P}}L\nonumber \\
 & =\sum_{k=1}^{n}\left[\frac{\nabla_{\boldsymbol{P}}\otimes\nabla_{\boldsymbol{P}}\gamma_{k}}{1+\gamma_{k}}-\left(\frac{\nabla_{\boldsymbol{P}}\gamma_{k}}{1+\gamma_{k}}\right)\otimes\left(\frac{\nabla_{\boldsymbol{P}}\gamma_{k}}{1+\gamma_{k}}\right)\right]\nonumber \\
 & =\sum_{k=1}^{n}\frac{-\bm{{\lambda}}_{k}\otimes\bm{{\lambda}}_{k}}{\left(1+\boldsymbol{P}\cdot\bm{{\lambda}}_{k}\right)^{2}}.\label{eq:H}
\end{align}
\end{linenomath}where again the second equality is derived using
eq.~(\ref{eq:gamma}) and $\nabla_{\boldsymbol{P}}\otimes\nabla_{\boldsymbol{P}}\gamma=0$.
The MLE uncertainties are~\cite{fisher1925}
\begin{equation}
\sigma_{P_{i}}=-\left[\left.H\right|_{\nabla_{\boldsymbol{P}}L=0}^{-1}\right]_{ii}^{1/2}\,\forall\,i\in\left\{ x,y,z\right\} .\label{eq:sigma_P}
\end{equation}

In standard analysis methods~\cite{A1RecoilPolarization2007,doria_thesis,Doria2015,Samo2017},
eq.~(\ref{eq:grad}) is solved numerically, with arbitrary initial
guess solutions. Given the complexity of the function and the fact
that in general there are no obvious good preliminary solution candidates
for the numerical process, this calculation can be CPU intensive,
and might converge to a local rather than the the global maximum.

\subsubsection{Analytical approximation\label{subsec:Analytical-approximation}}

In order to shorten the numerical process, and ensure convergence
to the correct solutions, following~\cite{BESSET1979}, we utilize
the property $\left|\gamma_{k}\right|<1$, and approximate the log-likelihood
function, expanding it in $\gamma_{k}$ up to $\mathcal{O}\left(\gamma_{k}^{3}\right)$:\begin{linenomath}
\begin{align}
L & \simeq\sum_{k=1}^{n}\left(\gamma_{k}-\frac{\gamma_{k}^{2}}{2}\right)\nonumber \\
 & =\sum_{k=1}^{n}\left(\boldsymbol{P}\cdot\bm{{\lambda}}_{k}-\frac{\left(\boldsymbol{P}\cdot\bm{{\lambda}}_{k}\right)^{2}}{2}\right)\nonumber \\
 & =\boldsymbol{P}\cdot\boldsymbol{b}-\boldsymbol{P}^{T}\mathsf{J}\boldsymbol{P}/2,\label{eq:linLike}
\end{align}
\end{linenomath}where\begin{linenomath}
\begin{align}
b_{i} & \equiv\sum_{k=1}^{n}\lambda_{i,k},\nonumber \\
J_{ij} & \equiv\sum_{k=1}^{n}\lambda_{i,k}\lambda_{j,k},\nonumber \\
i,j & \in\left\{ x,y,z\right\} .\label{eq:bI_def}
\end{align}
\end{linenomath} 

As in the previous section the connection between $\gamma$, $\boldsymbol{P}$
and $\bm{{\lambda}}$ is given by eq.~(\ref{eq:gamma}). Note that
the log-likelihood finite expansion approximation is performed only
in order to generate a proper set of initial solution candidates,
expedite the numerical process, and ensure that a correct final solution
is found to maximize $L$. Therefore, this approximation can be performed
even in cases where $\gamma_{k}$ is not much smaller than unity. 

The first MLEs solutions are obtained by equating the gradient of
the approximated log-likelihood function (eq.~(\ref{eq:linLike}))
to zero. The approximate simple algebraic form leads to a linear equation:
\begin{equation}
\nabla_{\boldsymbol{P}}L\simeq\boldsymbol{b}-\mathsf{J}\boldsymbol{P}=0,
\end{equation}
and thus, the first MLE candidates ($\boldsymbol{P}_{0}$) are given
by:
\begin{equation}
\boldsymbol{P}_{0}=\mathsf{J}^{-1}\boldsymbol{b}.\label{eq:P0}
\end{equation}
We use $\boldsymbol{P}_{0}$ to start the numerical process of Newton\textendash Raphson
steps, defined by: 
\begin{equation}
\boldsymbol{P}_{i}=\boldsymbol{P}_{i-1}-\left[H^{-1}\nabla_{\boldsymbol{P}}L\right]_{\boldsymbol{P}=\boldsymbol{P}_{i-1}}.\label{eq:NewtonStep}
\end{equation}

For $i=1$, we insert $\boldsymbol{P}_{0}$, $H^{-1}$ and $\nabla_{\boldsymbol{P}}L$
from eqs.~(\ref{eq:P0}), (\ref{eq:H}) and (\ref{eq:grad}), respectively,
into eq.~(\ref{eq:NewtonStep}), and obtain:\begin{linenomath}
\begin{equation}
\boldsymbol{P}_{1}=\boldsymbol{P}_{0}+\mathsf{J}'^{-1}\boldsymbol{b}',\label{eq:P1}
\end{equation}
\end{linenomath}where $\boldsymbol{b}'$ and $\mathsf{J}'$ are defined
by\begin{linenomath}
\begin{align}
b_{i}' & \equiv\sum_{k=1}^{n}\frac{\lambda_{i,k}}{1+\boldsymbol{P}_{0}\cdot\bm{{\lambda}}_{k}},\nonumber \\
J_{ij}' & \equiv\sum_{k=1}^{n}\frac{\lambda_{i,k}\lambda_{j,k}}{\left(1+\boldsymbol{P}_{0}\cdot\bm{{\lambda}}_{k}\right)^{2}},\nonumber \\
i,j & \in\left\{ x,y,z\right\} .\label{eq:bi_prime}
\end{align}
\end{linenomath}

$\boldsymbol{P}_{0}$ and one Newton\textendash Raphson step are good
estimates, and in practice, one step is enough to reach the absolute
maximum of $L$, limited only by the computer\textquoteright s precision.
In such a case the final MLEs are given by $\boldsymbol{P}_{1}$.
The MLEs\textquoteright{} covariance matrix and uncertainties are
given by the inverse of minus the likelihood\textquoteright s Hessian
(eq.~(\ref{eq:H})) and eq.~(\ref{eq:sigma_P}), respectively, both
calculated at $\boldsymbol{P}=\boldsymbol{P}_{1}$.

\subsection{The choice of a coordinate system\label{subsec:coordinate-system}}

\subsubsection{Motivation for using spherical coordinates\label{subsec:Motivation-spherical-coordinates}}

The above formalism is usually used in the standard Cartesian coordinates.
Separate polarization components can be extracted in this way with
unbiased and normally distributed MLEs, with reasonable uncertainties.

Nevertheless, many polarization-transfer experiments focus on the
components ratio $R\equiv P_{x}/P_{z}$ due to technical benefit and
physics interest~\cite{guy_ron_thesis}. The ratio cancels out the
inherent systematic uncertainty in beam polarization and analyzing
power, thus enabling the determination of $R$ to a better precision
than $P_{x}$ or $P_{z}$ separately. Furthermore, in the one photon
exchange approximation, the elastic electron proton scattering $R$
is proportional to the ratio of the elastic electric $G_{E}\left(Q^{2}\right)$
to the magnetic $G_{M}\left(Q^{2}\right)$ form factors at a given
four-momentum transfer $Q^{2}$~\cite{guy_ron_thesis}: \begin{linenomath}
\begin{equation}
R_{^{1}\text{H}}\equiv\left(\frac{P_{x}}{P_{z}}\right)_{^{1}\mathrm{H}}\!\!\!=-\frac{2M_{p}}{(E+E^{\prime})\tan(\theta_{e}/2)}\cdot\frac{G_{E}^{p}(Q^{2})}{G_{M}^{p}(Q^{2})},
\end{equation}
\end{linenomath} where $E$ ($E'$) is the incident (scattered) electron
energy, $\theta_{e}$ is the electron scattering angle, and $M_{p}$
is the proton mass.

As opposed to the separate components, the MLE of $R$ is biased,
and its distribution is skewed and has fat tails. Moreover, the uncertainty
in $R$ is propagated from the relative uncertainties and covariance
of $P_{x}$ and $P_{z}$, and is not derived directly from the likelihood~\cite{statistics_for_physicist}:
\begin{equation}
\Delta R\equiv R\sqrt{\Delta P_{x}^{2}/P_{x}^{2}+\Delta P_{z}^{2}/P_{z}^{2}-2\,\mathrm{Cov}\left[P_{x},P_{z}\right]/P_{x}P_{z}}.\label{eq:dR}
\end{equation}
The approximation behind the propagation assumes that the relative
uncertainties are small. Therefore, in small samples where the relative
uncertainties are large, the propagation approximation is invalid
and the formula is erroneous~\cite{fisher1925,statistics_for_physicist}.
This influences the minimal possible bin width. 

To circumvent these problems, we propose to analyze the polarization
in spherical coordinates, defined by: \begin{linenomath}
\begin{equation}
\boldsymbol{S}=\left(P,\vartheta,\varphi\right),\label{eq:S}
\end{equation}
\end{linenomath}where \begin{linenomath}
\begin{align}
P^{2} & \equiv P_{x}^{2}+P_{y}^{2}+P_{z}^{2},\nonumber \\
\rho^{2} & \equiv P_{y}^{2}+P_{z}^{2},\nonumber \\
\tan\vartheta & \equiv\rho/P_{x},\nonumber \\
\tan\varphi & \equiv P_{y}/P_{z}.\label{eq:spheric}
\end{align}
\end{linenomath}

Using spherical coordinates has several advantages. First, the magnitude
of the polarization vector can be extracted with higher accuracy than
each of its components. Second, two independent components, the polar
and azimuthal angles, do not depend on beam polarization. Third, since
the polarimeter can measure only two components, in some experiments
the spin precession is in the $yz$ plane so that $P_{z}$ and $P_{y}$
are mixed and $P_{x}$ remains unaffected by the rotation~\cite{pospischil_thesis}.
This increases the uncertainty in the measurement of $R$ as well.
In the new spherical coordinate system only the azimuthal angle ($\varphi$)
is affected by the mixing. Finally, by its definition, the cotangent
of the polar angle ($\vartheta$) is very similar to $R$, since in
elastic electron proton scattering $P_{y}=0$. Therefore, $\tau\equiv\cot\vartheta$
can be related to $G_{E}/G_{M}$ in a way similar to $R$. Because
$\vartheta$ is derived directly, its MLE distribution is unbiased
and almost normal, and its relative uncertainty is lower with respect
to that of $R$.

\subsubsection{The formalism in spherical coordinates }

The MLE in spherical coordinates is defined as $\boldsymbol{S}_{1}\equiv\boldsymbol{S}\left(\boldsymbol{P}_{1}\right)$,
where $\boldsymbol{S}\left(\boldsymbol{P}_{1}\right)$ is derived
explicitly by the transformation given in eq.~(\ref{eq:spheric})
at the value \textbf{$\boldsymbol{P}_{1}$ }(eq.~(\ref{eq:P1})).

In principle, the covariance matrix and the uncertainties of $\boldsymbol{S}_{1}$
could be derived by propagating the Cartesian uncertainties (eqs.~(\ref{eq:H}\textendash \ref{eq:sigma_P})
evaluated at $\boldsymbol{P}_{1}$) through the transformation of
eq.~(\ref{eq:spheric}). However, this would result in large uncertainties,
losing the main advantages of using spherical coordinates. We therefore
derive these values directly from the polarization components in spherical
coordinates, as elaborated below.

The log-likelihood function (eq.~(\ref{eq:logLike})) in spherical
coordinates is given by:\begin{linenomath}
\begin{align}
L\left(\boldsymbol{S}\right) & =\sum_{k=1}^{n}\ln\left(1+\gamma_{k}\right)\nonumber \\
 & =\sum_{k=1}^{n}\ln\left(1+\lambda_{x,k}P\cos\vartheta+\lambda_{y,k}P\sin\vartheta\sin\varphi+\lambda_{z,k}P\sin\vartheta\cos\varphi\right).
\end{align}
\end{linenomath}

As a preliminary technical step towards calculating the gradients
of the log-likelihood function, for non-vanishing polarization, we
first calculate the first derivatives of $\gamma$:\begin{linenomath}
\begin{align}
\partial_{P}\gamma & =\lambda_{x}\cos\vartheta+\lambda_{y}\sin\vartheta\sin\varphi+\lambda_{z}\sin\vartheta\cos\varphi\nonumber \\
 & =\gamma/P,\nonumber \\
\partial_{\vartheta}\gamma & =-\lambda_{x}P\sin\vartheta+\lambda_{y}P\cos\vartheta\sin\varphi+\lambda_{z}P\cos\vartheta\cos\varphi\nonumber \\
 & =-\lambda_{x}\rho+\xi\tau,\nonumber \\
\partial_{\varphi}\gamma & =\lambda_{y}P\sin\vartheta\cos\varphi-\lambda_{z}P\sin\vartheta\sin\varphi\nonumber \\
 & =\bar{\xi},
\end{align}
\end{linenomath}where $\tau=\cot\vartheta$, $\xi\equiv P\sin\vartheta\left(\lambda_{y}\sin\varphi+\lambda_{z}\cos\varphi\right)=\lambda_{y}P_{y}+\lambda_{z}P_{z}$,
and $\bar{\xi}\equiv P\sin\vartheta\left(\lambda_{y}\cos\varphi-\lambda_{z}\sin\varphi\right)=\lambda_{y}P_{z}-\lambda_{z}P_{y}$.
These can be combined to\begin{linenomath}
\begin{equation}
\nabla_{\boldsymbol{S}}\;\gamma=\begin{pmatrix}\gamma/P\\
-\lambda_{x}\rho+\xi\tau\\
\bar{\xi}
\end{pmatrix}.\label{eq:grad_s_g}
\end{equation}
\end{linenomath}The second derivatives of $\gamma$ are \begin{linenomath}
\begin{align}
\partial_{P}^{2}\gamma & =0,\nonumber \\
\partial_{\vartheta}\partial_{P}\gamma & =-\lambda_{x}\sin\vartheta+\left(\lambda_{y}\sin\varphi+\lambda_{z}\cos\varphi\right)\cos\vartheta\nonumber \\
 & =\left(\xi\tau-\lambda_{x}\rho\right)/P,\nonumber \\
\partial_{\varphi}\partial_{P}\gamma & =\left(\lambda_{y}\cos\varphi-\lambda_{z}\sin\varphi\right)\sin\vartheta\nonumber \\
 & =\bar{\xi}/P,\nonumber \\
\partial_{\vartheta}^{2}\gamma & =-P\left(\lambda_{x}\cos\vartheta+\left(\lambda_{y}\sin\varphi+\lambda_{z}\cos\varphi\right)\sin\vartheta\right)\nonumber \\
 & =-\gamma,\nonumber \\
\partial_{\varphi}\partial_{\vartheta}\gamma & =\left(\lambda_{y}\cos\varphi-\lambda_{z}\sin\varphi\right)P\cos\vartheta\nonumber \\
 & =\bar{\xi}\tau,\nonumber \\
\partial_{\varphi}^{2}\gamma & =\left(\lambda_{y}\sin\varphi+\lambda_{z}\cos\varphi\right)P\sin\vartheta\nonumber \\
 & =-\xi,
\end{align}
\end{linenomath}hence\begin{linenomath}
\begin{equation}
\nabla_{\boldsymbol{S}}\otimes\nabla_{\boldsymbol{S}}\;\gamma=\begin{pmatrix}\partial_{P}^{2}\gamma & \partial_{\vartheta}\partial_{P}\gamma & \partial_{\varphi}\partial_{P}\gamma\\
\partial_{\vartheta}\partial_{P}\gamma & \partial_{\vartheta}^{2}\gamma & \partial_{\varphi}\partial_{\vartheta}\gamma\\
\partial_{\varphi}\partial_{P}\gamma & \partial_{\varphi}\partial_{\vartheta}\gamma & \partial_{\varphi}^{2}\gamma
\end{pmatrix}=-\begin{pmatrix}0 & \frac{\lambda_{x}\rho-\xi\tau}{P} & -\frac{\bar{\xi}}{P}\\
\frac{\lambda_{x}\rho-\xi\tau}{P} & \gamma & -\bar{\xi}\tau\\
-\frac{\bar{\xi}}{P} & -\bar{\xi}\tau & \xi
\end{pmatrix}.\label{eq:H_s_g_comp}
\end{equation}
\end{linenomath}

We further define $\mathsf{J}''$:
\begin{equation}
\mathsf{J}''\equiv-\nabla_{\boldsymbol{S}}\otimes\nabla_{\boldsymbol{S}}L=\sum_{k=1}^{n}\left[\left(\frac{\nabla_{\boldsymbol{S}}\;\gamma_{k}}{1+\gamma_{k}}\right)\otimes\left(\frac{\nabla_{\boldsymbol{S}}\;\gamma_{k}}{1+\gamma_{k}}\right)-\frac{\nabla_{\boldsymbol{S}}\otimes\nabla_{\boldsymbol{S}}\;\gamma_{k}}{1+\gamma_{k}}\right].\label{eq:I_pp_1}
\end{equation}

The estimated asymptotic covariance matrix is the inverse of $\mathsf{J}''$,
and the uncertainties are given by the $\boldsymbol{S}$ equivalent
of eq.~(\ref{eq:sigma_P}), both calculated at $\boldsymbol{S}_{1}$.

\subsection{Varying polarization\label{subsec:formalism_for_varying_polarization}}

In the previous sections, we assumed that the polarization $\boldsymbol{P}$
is constant for all of the events. In reality the polarization of
each event is different, and depends on several physical variables,
such as momentum transfer squared ($Q^{2}$), missing momentum, virtuality~\cite{deep2012PLB,ceepLet},
etc. Usually one sets up an experiment in a way that all but one of
the variables are fixed, and investigates the polarization dependence
on that one variable. In the following, we present our tool for extracting
the polarization in this case of non-constant polarization.

In the most general case, the polarization $\boldsymbol{P}$ has a
different value for each event, and then log-likelihood $L$ takes
the form\begin{linenomath}
\begin{equation}
L=\sum_{k=1}^{n}\ln\left(1+\boldsymbol{P}_{k}\cdot\bm{{\lambda}}_{k}\right).
\end{equation}
\end{linenomath}In this case, $L$ has $3n$ parameters, where $n$
is the number of events, so that the problem of extracting $\boldsymbol{P}$
is under-determined. However, if the polarization depends on a certain
measured variable $v$, the derivation of $\boldsymbol{P}\left(v\right)$
is possible by maximizing a generalized likelihood function, as described
below.

\subsubsection{Binned likelihood \label{subsec:Binned-likelihood}}

The traditional way of describing the dependence of polarization data
(or any other data) on a variable, is by displaying the measurement
in bins of that variable. The events are divided into $N+1$ bins\footnote{We chose to divide to $N+1$ bins rather than $N$, for consistency
of the algebra between the binned solution and the following continuous
description.} of the variable $v$, where each bin is defined by the range $\left(v_{l-1},v_{l}\right)$.
Binning is expressed mathematically by assigning a weight $w_{lk}$
to each event $k$, which takes the value $1$ when $v$ is in the
bin range $\left(v_{l-1},v_{l}\right)$ and $0$ otherwise, namely,
$w_{lk}=\begin{cases}
1 & v_{k}\in\left(v_{l-1},v_{l}\right)\\
0 & v_{k}\notin\left(v_{l-1},v_{l}\right)
\end{cases}$.

The likelihood function now obtains a generalized form that includes
the polarization binning in $v$, $\boldsymbol{P}\left(v\right)=\sum_{l=0}^{N}w_{l}\left(v\right)\,\boldsymbol{P}_{l}$,
where $\boldsymbol{P}_{l}$ are the $3\left(N+1\right)$ parameters
of $\boldsymbol{P}\left(v\right)$, and $w_{l}\left(v\right)$ are
the weight functions that satisfy $w_{lk}=w_{l}\left(v_{k}\right)$.

$L$ is now given by: \begin{linenomath}
\begin{equation}
L=\sum_{k=1}^{n}\ln\left(1+\sum_{l=0}^{N}w_{lk}\boldsymbol{P}_{l}\cdot\bm{{\lambda}}_{k}\right).\label{eq:loglikeWght}
\end{equation}
\end{linenomath}The MLE of $\boldsymbol{P}_{l}$ and their uncertainties
can be found by extending the definitions of $\boldsymbol{b}$, $\mathsf{J}$,
$\boldsymbol{b}'$ and $\mathsf{J}'$ from eqs.~(\ref{eq:bI_def})
and (\ref{eq:bi_prime}) respectively:\begin{linenomath}
\begin{align}
b_{3l+i} & \equiv\sum_{k=1}^{n}w_{lk}\lambda_{i,k},\nonumber \\
J_{3l+i,3l'+j} & \equiv\sum_{k=1}^{n}w_{lk}w_{l'k}\lambda_{i,k}\lambda_{j,k},\nonumber \\
b_{3l+i}' & \equiv\sum_{k=1}^{n}\frac{w_{lk}\lambda_{i,k}}{1+\gamma_{k}},\nonumber \\
J_{3l+i,3l'+j}' & \equiv\sum_{k=1}^{n}\frac{w_{lk}w_{l'k}\lambda_{i,k}\lambda_{j,k}}{\left(1+\gamma_{k}\right)^{2}},\nonumber \\
i,j & \in\left\{ x=1,y=2,z=3\right\} ,\label{eq:bI_cor}
\end{align}
\end{linenomath} where $\gamma_{k}\equiv\sum_{l=0}^{N}w_{lk}\boldsymbol{P}_{l}\cdot\bm{{\lambda}}_{k}$
extends eq.~(\ref{eq:gamma}).

The algebra of eqs.~(\ref{eq:P0}), (\ref{eq:P1}) and (\ref{eq:sigma_P})
still holds:\begin{linenomath}
\begin{align*}
\boldsymbol{P}_{0} & =\mathsf{J}^{-1}\boldsymbol{b}\\
\boldsymbol{P}_{1} & =\boldsymbol{P}_{0}+\mathsf{J}'^{-1}\boldsymbol{b}'\\
\sigma_{P_{1,i}} & =-\left[\left.\mathsf{J}'\right|_{\boldsymbol{P}_{1}}^{-1}\right]_{ii}^{1/2}\,\forall\,i\in\left\{ x,y,z\right\} .
\end{align*}
\end{linenomath}Since in the binned likelihood case, the weights
are merely indications of whether the event is in a specific bin or
not, and the parameters $\boldsymbol{P}_{l}$ ($l=0$ to $N$) are
the values of $\boldsymbol{P}$ in each of the $N+1$ bins, the matrix
$\mathsf{J}$ (eq.~(\ref{eq:bI_cor})) now becomes a block matrix,
where each block corresponds to a bin. Since by definition the polarization
is constant for each bin, the results for $\boldsymbol{P}_{1}$ can
also be extracted from each block of the matrix according to the formalism
described in \subsecref{Analytical-approximation}.

As discussed in \secref{Introduction}, using bins is a compromise
regarding the amount of physical information that may be extracted
from the measurement. In the following, we present a method that also
accounts for the location of each event inside the bin.

\subsubsection{Piecewise continuous linear dependence}

The first step of generalizing the binned approach is to assume a
continuous linear dependence of the polarization within each bin\footnote{Of now $N$ bins, rather than $N+1$ in the previous section.}.
The polarization is thus defined as a piecewise continuous function
$\boldsymbol{P}\left(v\right)=\boldsymbol{p}_{l}\left(v\right)\,\forall\,v\in\left(v_{l-1},v_{l}\right)$,
where $\boldsymbol{p}_{l}\left(v\right)$ is a linear interpolation
between the polarization at the edges, $\boldsymbol{P}_{l-1}$ and
$\boldsymbol{P}_{l}$:
\begin{equation}
\boldsymbol{p}_{l}\left(v\right)=\frac{v_{l}-v}{v_{l}-v_{l-1}}\boldsymbol{P}_{l-1}+\frac{v-v_{l-1}}{v_{l}-v_{l-1}}\boldsymbol{P}_{l}=\left(1-x\right)\boldsymbol{P}_{l-1}+x\boldsymbol{P}_{l}
\end{equation}
$v_{l}$ and $\boldsymbol{P}_{l}$ are now the values at the edges
of the a-priori defined bins, and 
\begin{equation}
x\left(v\right)\equiv\frac{v-v_{l-1}}{v_{l}-v_{l-1}}\;\forall\;v\in\left(v_{l-1},v_{l}\right)\label{eq:x(v)}
\end{equation}
 is the relative position in the bin.

The weights are now continuous and linear inside the relevant bin
and zero outside: 
\begin{equation}
w_{lk}=\begin{cases}
1-x & v_{k}\in\left(v_{l-1},v_{l}\right)\\
x & v_{k}\in\left(v_{l},v_{l+1}\right)\\
0 & v_{k}\notin\left(v_{l-1},v_{l+1}\right)
\end{cases}.\label{eq:w_lin}
\end{equation}
These weights are then inserted into eq.~(\ref{eq:bI_cor}) to deduce
the MLE values and their uncertainties.

\subsubsection{Cubic spline interpolation weighting\label{subsec:Cubic-spline-interpolation}}

The behavior of the measured polarization may not be limited to linear
form. However, high order polynomial interpolation is susceptible
to Runge\textquoteright s phenomenon of oscillation at the edges of
the interval. Therefore, we opt to perform spline interpolation.

In interpolating problems, spline interpolation is often preferred
over polynomial interpolation because it yields similar results, even
when using lower degree polynomials. We outline here a derivation
of the polarization via maximizing the likelihood for the commonly
used cubic spline. 

A cubic spline $\boldsymbol{P}\left(v\right)$ with $N+1$ knots is
a piecewise function constructed of $N$ cubic functions 
\begin{equation}
\boldsymbol{p}_{l}\left(v\right)=\sum_{i=0}^{3}\tilde{\boldsymbol{a}}_{il}v^{i}=\sum_{i=0}^{3}\boldsymbol{a}_{il}x^{i},\label{eq:Pl}
\end{equation}
where $x$ defines the relative position between the knots (that is,
the range within the bin) as in eq.~(\ref{eq:x(v)}).

The parameters are the values at the knots $\boldsymbol{P}\left(v_{l}\right)=\boldsymbol{P}_{l}$.
We demand $C^{2}$ continuity:\begin{linenomath}\begin{subequations}
\begin{align}
\boldsymbol{p}_{l}\left(v_{l}\right) & =\boldsymbol{p}_{l+1}\left(v_{l}\right),\label{eq:spline_cont:0}\\
\boldsymbol{p}_{l}'\left(v_{l}\right) & =\boldsymbol{p}_{l+1}'\left(v_{l}\right),\label{eq:spline_cont:1}\\
\boldsymbol{p}_{l}''\left(v_{l}\right) & =\boldsymbol{p}_{l+1}''\left(v_{l}\right).\label{eq:spline_cont:2}
\end{align}
\end{subequations}\end{linenomath}We note that from here on the
derivation is given for one dimension, since the weight is defined
per event, so each polarization component has the same weight. Generalizing
to three dimensions merely means writing the same equations three
times with different component indices.

We now proceed to write explicitly all the constraints on the cubic
spline coefficients $a_{il}$. We start by writing the expressions
for $p_{l}$ at the $N+1$ knots. The first $N$ equations are for
the left edges of the respective bins ($x=0$). Inserting $x=0$ into
eq.~(\ref{eq:Pl}) one obtains:\begin{linenomath} 
\begin{equation}
p_{l}\left(v_{l-1}\right)=a_{0l}=P_{l-1}\;\;\forall\;l\in\left(1,N\right).
\end{equation}
\end{linenomath}The ($N+1$)th equation is defined at the last knot,
namely the right edge ($x=1$) of the rightmost bin. Eq.~(\ref{eq:Pl})
becomes: \begin{linenomath}
\begin{equation}
p_{N}\left(v_{N}\right)=\sum_{i=0}^{3}a_{i,N}=P_{N}.
\end{equation}
\end{linenomath}We can combine these $N+1$ equations into the matrix
equation $\mathsf{A}_{P}\vec{a}=\vec{P}$, where\begin{linenomath}
\begin{align}
\mathsf{A}_{P} & \equiv\left(\begin{array}{ccccc|ccc|ccc|ccc}
1 & 0 & \cdots & 0 & 0 & 0 & \cdots & 0 & 0 & \cdots & 0 & 0 & \cdots & 0\\
0 & 1 & \cdots & 0 & 0 & 0 & \cdots & 0 & 0 & \cdots & 0 & 0 & \cdots & 0\\
\vdots & \vdots & \ddots & \vdots & \vdots & \vdots & \ddots & \vdots & \vdots & \ddots & \vdots & \vdots & \ddots & \vdots\\
0 & 0 & \cdots & 1 & 0 & 0 & \cdots & 0 & 0 & \cdots & 0 & 0 & \cdots & 0\\
0 & 0 & \cdots & 0 & 1 & 0 & \cdots & 0 & 0 & \cdots & 0 & 0 & \cdots & 0\\
0 & 0 & \cdots & 0 & 1 & 0 & \cdots & 1 & 0 & \cdots & 1 & 0 & \cdots & 1
\end{array}\right),\nonumber \\
\vec{a}^{T} & \equiv\left(\begin{array}{ccc|ccc|ccc|ccc}
a_{0,1} & \cdots & a_{0,N} & a_{1,1} & \cdots & a_{1,N} & a_{2,1} & \cdots & a_{2,N} & a_{3,1} & \cdots & a_{3,N}\end{array}\right),\nonumber \\
\vec{P}^{T} & \equiv\begin{pmatrix}P_{0} & \cdots & P_{N}\end{pmatrix}.
\end{align}
\end{linenomath}The first order of continuity (eq.~(\ref{eq:spline_cont:0}))
is defined at each of the $N-1$ interior knots ($l\in\left(1,N-1\right)$).
When inserting eq.~(\ref{eq:Pl}) into eq.~(\ref{eq:spline_cont:0}),
the left hand side of the equation gives $x=1$ at the left side of
the knot, and the right hand side gives $x=0$ at the right side.
Thus, the first order of continuity takes the form\begin{linenomath}
\begin{equation}
a_{0,l}+a_{1,l}+a_{2,l}+a_{3,l}-a_{0,l+1}=0\;\;\forall\;l\in\left(1,N-1\right),
\end{equation}
\end{linenomath}and can be written in matrix form as $\mathsf{A}_{0}\vec{a}=\vec{0}$,
where \begin{linenomath}
\begin{equation}
\mathsf{A}_{0}\equiv\left(\begin{array}{cccccc|ccccc|ccccc|ccccc}
1 & -1 & 0 & \cdots & 0 & 0 & 1 & 0 & \cdots & 0 & 0 & 1 & 0 & \cdots & 0 & 0 & 1 & 0 & \cdots & 0 & 0\\
0 & 1 & -1 & \cdots & 0 & 0 & 0 & 1 & \cdots & 0 & 0 & 0 & 1 & \cdots & 0 & 0 & 0 & 1 & \cdots & 0 & 0\\
\vdots & \vdots & \vdots & \ddots & \vdots & \vdots & \vdots & \vdots & \ddots & \vdots & \vdots & \vdots & \vdots & \ddots & \vdots & \vdots & \vdots & \vdots & \ddots & \vdots & \vdots\\
0 & 0 & 0 & \cdots & 1 & -1 & 0 & 0 & \cdots & 1 & 0 & 0 & 0 & \cdots & 1 & 0 & 0 & 0 & \cdots & 1 & 0
\end{array}\right).
\end{equation}

\end{linenomath}In a similar way, the first order of smoothness (eq.~(\ref{eq:spline_cont:1}))
is \begin{linenomath}
\begin{equation}
a_{1,l}+2a_{2,l}+3a_{3,l}-\tilde{r}_{l}a_{1,l+1}=0\;\;\forall\;l\in\left(1,N-1\right),
\end{equation}
 \end{linenomath}where the term $\tilde{r}_{l}\equiv\frac{v_{l}-v_{l-1}}{v_{l+1}-v_{l}}$
results from the change of variables in eq.~(\ref{eq:x(v)}). This
can be written in matrix form as $\mathsf{A}_{1}\vec{a}=\vec{0}$,
where \begin{linenomath}
\begin{equation}
\mathsf{A}_{1}\equiv\left(\begin{array}{ccc|cccccc|ccccc|ccccc}
0 & \cdots & 0 & 1 & -\tilde{r}_{1} & 0 & \cdots & 0 & 0 & 2 & 0 & \cdots & 0 & 0 & 3 & 0 & \cdots & 0 & 0\\
0 & \cdots & 0 & 0 & 1 & -\tilde{r}_{2} & \cdots & 0 & 0 & 0 & 2 & \cdots & 0 & 0 & 0 & 3 & \cdots & 0 & 0\\
\vdots & \ddots & \vdots & \vdots & \vdots & \vdots & \ddots & \vdots & \vdots & \vdots & \vdots & \ddots & \vdots & \vdots & \vdots & \vdots & \ddots & \vdots & \vdots\\
0 & \cdots & 0 & 0 & 0 & 0 & \cdots & 1 & -\tilde{r}_{N-1} & 0 & 0 & \cdots & 2 & 0 & 0 & 0 & \cdots & 3 & 0
\end{array}\right).
\end{equation}

\end{linenomath}Lastly, the second order of smoothness (eq.~(\ref{eq:spline_cont:2}))
is \begin{linenomath}
\begin{equation}
a_{2,l}+3a_{3,l}-\tilde{r}_{l}^{2}a_{2,l+1}=0\;\;\forall\;l\in\left(1,N-1\right),
\end{equation}
\end{linenomath}which can be written as $\mathsf{A}_{2}\vec{a}=\vec{0}$,
where \begin{linenomath}
\begin{equation}
\mathsf{A}_{2}\equiv\left(\begin{array}{ccc|ccc|cccccc|ccccc}
0 & \cdots & 0 & 0 & \cdots & 0 & 1 & -\tilde{r}_{1}^{2} & 0 & \cdots & 0 & 0 & 3 & 0 & \cdots & 0 & 0\\
0 & \cdots & 0 & 0 & \cdots & 0 & 0 & 1 & -\tilde{r}_{2}^{2} & \cdots & 0 & 0 & 0 & 3 & \cdots & 0 & 0\\
\vdots & \ddots & \vdots & \vdots & \ddots & \vdots & \vdots & \vdots & \vdots & \ddots & \vdots & \vdots & \vdots & \vdots & \ddots & \vdots & \vdots\\
0 & \cdots & 0 & 0 & \cdots & 0 & 0 & 0 & 0 & \cdots & 1 & -\tilde{r}_{N-1}^{2} & 0 & 0 & \cdots & 3 & 0
\end{array}\right).
\end{equation}

\end{linenomath}The above constitute a total of $4N-2$ equations.
In order to obtain a closed solution for $w_{l}\left(v\right)$ we
need two more equations (constraints). These are obtained by imposing
boundary conditions at the two edges of the full $v$ range. Common
boundary conditions are parabolic ($x^{2}$, instead of cubic) dependence
at the first and last knots, namely $a_{31}=0$ and $a_{3N}=0$, or
equality of the third derivative in the second and next-to-last knots,
namely $a_{3,1}-\tilde{r}_{1}^{3}a_{3,2}=0$ and $a_{3,N-1}-\tilde{r}_{N-1}^{3}a_{3,N}=0$.
These can be written in matrix form as $\mathsf{A}_{b}\vec{a}=\vec{0}$
with \begin{linenomath}
\begin{equation}
\mathsf{A}_{b}^{1}\equiv\left(\begin{array}{ccc|ccc|ccc|ccccc}
0 & \cdots & 0 & 0 & \cdots & 0 & 0 & \cdots & 0 & 1 & 0 & \cdots & 0 & 0\\
0 & \cdots & 0 & 0 & \cdots & 0 & 0 & \cdots & 0 & 0 & 0 & \cdots & 0 & 1
\end{array}\right)
\end{equation}
\end{linenomath}or\begin{linenomath}
\begin{equation}
\mathsf{A}_{b}^{2}\equiv\left(\begin{array}{ccc|ccc|ccc|ccccccc}
0 & \cdots & 0 & 0 & \cdots & 0 & 0 & \cdots & 0 & 1 & -\tilde{r}_{1}^{3} & 0 & \cdots & 0 & 0 & 0\\
0 & \cdots & 0 & 0 & \cdots & 0 & 0 & \cdots & 0 & 0 & 0 & 0 & \cdots & 0 & 1 & -\tilde{r}_{N-1}^{3}
\end{array}\right),
\end{equation}
\end{linenomath}respectively.

The entire set of the \emph{above} $4N$ equations can be summarized
to\begin{linenomath}
\begin{equation}
\mathsf{A}\vec{a}\equiv\begin{pmatrix}\\
\mathsf{A}_{P}\\
\\
\text{\_\_\_}\\
\mathsf{A}_{0}\\
\mathsf{A}_{1}\\
\mathsf{A}_{2}\\
\mathsf{A}_{b}
\end{pmatrix}\vec{a}=\begin{pmatrix}P_{0}\\
\vdots\\
P_{n}\\
\text{\_\_\_}\\
0\\
\vdots\\
\vdots\\
0
\end{pmatrix}.\label{eq:Aa--p}
\end{equation}

\end{linenomath}Now, on the one hand, by definition, $P\left(v\right)=\sum_{l=0}^{N}w_{l}\left(v\right)P_{l}$.
On the other hand, $P\left(v\right)=\sum_{i=0}^{3}a_{il'}x^{i}\left(v\right)\;\;\forall\;v\in\left(v_{l'-1},v_{l'}\right)$.
We can compare both terms, to extract $w_{l}\left(v\right)$, through
the inversion of eq.~(\ref{eq:Aa--p}):\begin{linenomath}
\begin{align}
P\left(v\right) & =\sum_{i=0}^{3}a_{il'}x^{i}\left(v\right)\;\;\forall\;v\in\left(v_{l'-1},v_{l'}\right)\nonumber \\
 & =\sum_{i=0}^{3}\sum_{l=0}^{N}\mathsf{A}_{i\cdot N+l',l}^{-1}\,P_{l}\,x^{i}\left(v\right)\nonumber \\
 & =\sum_{l=0}^{N}w_{l}\left(v\right)P_{l}.\label{eq:explain-w}
\end{align}
\end{linenomath}Therefore, when solving for $P_{l}$, the weight
functions are: \begin{linenomath}
\begin{equation}
w_{l}\left(v\right)=\sum_{i=0}^{3}\mathsf{A}_{i\cdot N+l',l}^{-1}\,x^{i}\left(v\right)\;\;\forall\;v\in\left(v_{l'-1},v_{l'}\right).\label{eq:cubic_spline_ws}
\end{equation}
\end{linenomath}The event weights $w_{lk}=w_{l}\left(v_{k}\right)$
are inserted into eq.~(\ref{eq:bI_cor}), which leads to the derivation
of the MLEs and their uncertainties.

One can extend this derivation for a natural spline of higher dimensionality
$D$, while keeping the same number of parameters, by demanding a
$C^{D-1}$ continuity at the knots. E.g., for a fifth level spline,
one should extend $\mathsf{A}_{0}$, $\mathsf{A}_{1}$, $\mathsf{A}_{2}$
and $\mathsf{A}_{b}$, and add the corresponding matrices $\mathsf{A}_{3}$
and $\mathsf{A}_{4}$ to eq.~(\ref{eq:Aa--p}).

\subsection{Comparison of the data to theoretical models and other measurements\label{subsec:Comparison-to-theoretical}}

It is customary to include theoretical models in plots depicting measured
data. Comparing the two is essential for extraction of the underling
physics. When comparing measured data with a theoretical model, one
usually attempts to test the hypothesis that the model describes the
data (e.g., by calculating the $p$-value). In more advanced comparison
efforts, one tries to quantify the difference between the model and
the measurement, thus deducing to what extent the data exhibits physical
effects that are not described by the model. 

In this section, we present methods that enable data and calculation
comparisons, which utilize the maximum information that is available
from the measurements and the associated theoretical model.

\subsubsection{Quantifying the difference between measurements and a theoretical
model\label{subsec:Quantification-of-the}}

The common method to compare measurements to a theoretical model is
to observe the overall trend of the data as a function of one or more
variables, and perform a global comparison to the trend predicted
by the model.

We introduce a method that focuses on observing the ratio between
the data and a theoretical model prediction for each measured event.
This avoids possible loss of information due to inherent assumptions
that are included in global trends. 

We define the ratio of a measurement to a model as $r_{i}\equiv P_{i}^{\mathrm{exp}}/P_{i}^{\mathrm{calc}}$,
where $\boldsymbol{P}^{\mathrm{exp}}$ is the measured polarization
in a specific event, and $\boldsymbol{P}^{\mathrm{calc}}$ is the
calculated theoretical model value for the kinematics of that specific
event. We then calculate the MLEs of $\boldsymbol{r}$ by maximizing
the following log-likelihood function: \begin{linenomath}
\begin{equation}
L\left(\boldsymbol{r}\right)=\sum_{k=1}^{n}\ln\left(1+\sum_{i\in\left\{ x,y,z\right\} }r_{i}P_{i,k}^{\mathrm{calc}}\lambda_{i,k}\right)\label{eq:L_r}
\end{equation}
\end{linenomath}By redefining $\lambda_{i,k}\rightarrow P_{i,k}^{\mathrm{calc}}\lambda_{i,k}$
the log-likelihood reverts to the form in eq.~(\ref{eq:logLike}),
and can be solved for $\boldsymbol{r}_{1}$ according to eq.~(\ref{eq:P1}). 

We can generalize this formalism to varying polarization that depends
on a certain variable $v$, and obtain the MLEs for $\boldsymbol{r}(v)$,
as detailed in \subsecref{formalism_for_varying_polarization}.

A special case of the above is when the only difference between the
measurements and the theoretical model is a scalar factor that is
common to all components, $r\equiv P^{\mathrm{exp}}/P^{\mathrm{\mathrm{calc}}}=P_{x}^{\mathrm{exp}}/P_{x}^{\mathrm{calc}}=P_{y}^{\mathrm{exp}}/P_{y}^{\mathrm{calc}}=P_{z}^{\mathrm{exp}}/P_{z}^{\mathrm{calc}}$.
An example is a polarization transfer measurement to a free proton,
as in the $p\left(\vec{e},e'\vec{p}'\right)$ reaction, where the
proton form factor ratio is well known, and the aim is to extract
the analyzing power, the beam polarization, or their product. If this
is the case, eq.~(\ref{eq:L_r}) takes the form\begin{linenomath}
\begin{equation}
L\left(r\right)=\sum_{k=1}^{n}\ln\left(1+r\boldsymbol{P}_{k}^{\mathrm{calc}}\cdot\bm{{\lambda}}_{k}\right).\label{eq:L_scalar}
\end{equation}
\end{linenomath}The MLE of $r$ and its uncertainties can be found
by adjusting the definitions of $\boldsymbol{b}$, $\mathsf{J}$,
$\boldsymbol{b}'$ and $\mathsf{J}'$ from eqs.~(\ref{eq:bI_def})
and (\ref{eq:bi_prime}) respectively:\begin{linenomath}
\begin{align}
b & \equiv\sum_{k=1}^{n}\boldsymbol{P}_{k}^{\mathrm{calc}}\cdot\bm{{\lambda}}_{k},\nonumber \\
J & \equiv\sum_{k=1}^{n}\left(\boldsymbol{P}_{k}^{\mathrm{calc}}\cdot\bm{{\lambda}}_{k}\right)^{2},\nonumber \\
b' & \equiv\sum_{k=1}^{n}\frac{\boldsymbol{P}_{k}^{\mathrm{calc}}\cdot\bm{{\lambda}}_{k}}{1+r\boldsymbol{P}_{k}^{\mathrm{calc}}\cdot\bm{{\lambda}}_{k}},\nonumber \\
J' & \equiv\sum_{k=1}^{n}\left(\frac{\boldsymbol{P}_{k}^{\mathrm{calc}}\cdot\bm{{\lambda}}_{k}}{1+r\boldsymbol{P}_{k}^{\mathrm{calc}}\cdot\bm{{\lambda}}_{k}}\right)^{2}.\label{eq:bI_scalar}
\end{align}
\end{linenomath}The algebra of eqs.~(\ref{eq:P0}), (\ref{eq:P1})
and (\ref{eq:sigma_P}) still holds:\begin{linenomath}
\begin{align}
r_{0} & =J^{-1}b,\nonumber \\
r_{1} & =r_{0}+\left[J'^{-1}b'\right]_{r=r_{0}},\nonumber \\
\sigma_{r_{1}} & =\left.J'\right|_{r=r_{1}}^{-1/2}.\label{eq:r01_sr}
\end{align}
\end{linenomath}

Notice that in this special case, even if $r$ is not expected to
depend on any kinematical variable, it could vary with time ($t$),
and thus be different for each event. In order to take such temporal
variations into account, one can generalize eqs.~(\ref{eq:L_scalar}\textendash \ref{eq:r01_sr})
to $r$ that depends on a single variable (time in this case) according
to \subsecref{formalism_for_varying_polarization}, and deduce $r\left(t\right)$.

\subsubsection{Tests of statistical consistency between measurements and calculations
($p$-value)\label{par:Testing-if-the}}

We define $L_{H_{1}}$ to be the log-likelihood function of the experimental
data, whose maximum is given by the MLEs $r_{i,1}\,\forall\,i\in\left\{ x,y,z\right\} $,
and $L_{H_{0}}$ to be the log-likelihood function of the theoretical
model, whose MLEs are $r_{i}=1\,\forall\,i\in\left\{ x,y,z\right\} $,
by definition. The \textquoteleft null hypothesis\textquoteright{}
($H_{0}$), which is the hypothesis that the experimental data and
the theoretical model are consistent, is tested by using the ratio
of these two likelihoods, or the difference of the two log-likelihoods,
$\Lambda\equiv L_{H_{1}}-L_{H_{0}}=L\left(\boldsymbol{r}_{1}\right)-L\left(\boldsymbol{1}\right)$,
where $L\left(\boldsymbol{r}\right)$ is given in eq.~(\ref{eq:L_r}). 

According to Wilks\textquoteright{} theorem for log-likelihood ratio
statistics, for a large number of events, $2\Lambda$ has a $\chi^{2}$
distribution with $\nu$ degrees of freedom, where $\nu$ is the difference
between the numbers of parameters in the compared models~\cite{fisher1925,statistics_for_physicist}.
In this case, one \textquoteleft model\textquoteright , the experimental
data, has 3 parameters ($r_{i,1}$), and the theoretical model (the
\textquoteleft null model\textquoteright ) has no parameters (since
$r_{i}=1$) so $\nu=3$. The \textquoteleft null hypothesis\textquoteright{}
test is performed by calculating the cumulative distribution function
(CDF) of the distribution, $p\equiv p\left(\chi^{2},\nu\right)=p\left(2\Lambda,3\right)$.
If $p>5\%$, we cannot reject the \textquoteleft null hypothesis\textquoteright ,
namely, the experiment and model are consistent~\cite{fisher1925}.

The above process can be generalized to varying polarization. If $\boldsymbol{r}=\boldsymbol{r}(v)$
and the $v$ range is divided into $N$ bins, then the number of degrees
of freedom changes to $\nu=3(N+1)$, and other than that there is
no change.

A more sophisticated test of the validity of the model is to check
consistency up to a normalization factor for the polarization-transfer,
$r$, which might arise from, e.g., a systematic error in the beam
polarization. The likelihood-ratio log is $\Lambda=L\left(\boldsymbol{r}_{1}\right)-L\left(r_{1}\right)$,
where $L\left(\boldsymbol{r}\right)$ and $L\left(r\right)$ are given
by eqs.~(\ref{eq:L_r}\textendash \ref{eq:L_scalar}), respectively.
Since the theoretical model has now one parameter, the difference
in degrees of freedom is 2, and the \textquoteleft null hypothesis\textquoteright{}
test is $p\left(2\Lambda,2\right)>5\%$.

Also here, a varying $\boldsymbol{r}\left(v\right)$ with $N$ bins
will increase the number of degrees of freedom to $2\left(N+1\right)$.

\subsubsection{Plotting model predictions with data}

When depicting experimental data, it is common to plot a model\textquoteright s
prediction with it. However, while the measurement result ($\boldsymbol{P}^{\mathrm{exp}}$)
is calculated for an ensemble of events, the model ($\boldsymbol{P}^{\mathrm{calc}}$)
is calculated for each event separately. 

The traditional way to present the model is to take its average over
the events that constitute the experimental sample, $\boldsymbol{P}_{a}^{\mathrm{calc}}\equiv n^{-1}\sum_{k=1}^{n}\boldsymbol{P}_{k}^{\mathrm{calc}}$.
However, one can perform a better comparison by using a weighted average
of the model dictated by the experimental sample. As will be evident
from the derivation below, this weight turns out to be the effective
analyzing power of the event. The experimental events are weighted
by the analyzing power, and now the theoretical ones are weighted
in the same manner.

When observing a plot with data points and the respective model predictions,
the quarry is the difference between the two: $\bm{{\delta}}\equiv\boldsymbol{P}^{\mathrm{exp}}-\boldsymbol{P}^{\mathrm{calc}}$.
To estimate $\bm{{\delta}}$ we calculate the log-likelihood:
\begin{equation}
L\left(\bm{{\delta}}\right)=\sum_{k=1}^{n}\ln\left(1+\left(\bm{{\delta}}+\boldsymbol{P}_{k}^{\mathrm{calc}}\right)\cdot\bm{{\lambda}}_{k}\right).
\end{equation}
Taking the linear approximation of \subsecref{Analytical-approximation},
we obtain
\begin{equation}
L\left(\bm{{\delta}}\right)\simeq\sum_{k=1}^{n}\left[\left(\bm{{\delta}}+\boldsymbol{P}_{k}^{\mathrm{calc}}\right)\cdot\bm{{\lambda}}_{k}-\frac{1}{2}\left(\left(\bm{{\delta}}+\boldsymbol{P}_{k}^{\mathrm{calc}}\right)\cdot\bm{{\lambda}}_{k}\right)^{2}\right],
\end{equation}
and using the definitions of eq.~(\ref{eq:bI_def}), this expression
is reduced to
\begin{equation}
L\left(\bm{{\delta}}\right)\simeq\bm{{\delta}}\cdot\boldsymbol{b}+C_{1}-\bm{{\delta}}^{T}\mathsf{J}\bm{{\delta}}/2-\bm{{\delta}}^{T}\mathsf{J}\boldsymbol{P}_{w}^{\mathrm{calc}}-C_{2},
\end{equation}
where $C_{1}$ and $C_{2}$ are constants that are independent of
$\bm{{\delta}}$, and 
\begin{equation}
\boldsymbol{P}_{w}^{\mathrm{calc}}\equiv\mathsf{J}^{-1}\sum_{k=1}^{n}\left(\left(\boldsymbol{P}_{k}^{\mathrm{calc}}\cdot\bm{{\lambda}}_{k}\right)\bm{{\lambda}}_{k}\right).
\end{equation}
 Comparing this to eq.~(\ref{eq:gamma}), one indeed sees that $\boldsymbol{P}_{w}^{\mathrm{calc}}$
is weighted by an effective analyzing power, as described above. 

The linear log-likelihood is maximized at 
\begin{equation}
\bm{{\delta}}_{0}=\mathsf{J}^{-1}\left(\boldsymbol{b}-\mathsf{J}\boldsymbol{P}_{w}^{\mathrm{calc}}\right)=\boldsymbol{P}_{0}-\boldsymbol{P}_{w}^{\mathrm{calc}}.
\end{equation}
The definition of $\boldsymbol{P}_{w}^{\mathrm{calc}}$ can be extended
to a continuous description of the data by redefining $\lambda_{i,k}\rightarrow w_{lk}\lambda_{i,k}$,
where $w_{lk}$ were defined in \subsecref{formalism_for_varying_polarization}.

\subsubsection{Comparisons between measured results from two experiments}

In addition to comparing experimental data to theoretical calculations,
there is interest in comparing two experimental data sets, to determine
if they are consistent. For example, one could be interested in comparing
polarization transfer to two different target nuclei. In such a comparison,
one needs to account for overall differences between the data sets,
for example, beam polarization.

We would like to test whether two data sets ($A$ and $B$) describe
the same polarization-transfer, i.e., if they differ from each other
by a scalar factor $r\equiv P^{B}/P^{A}=P_{x}^{B}/P_{x}^{A}=P_{y}^{B}/P_{y}^{A}=P_{z}^{B}/P_{z}^{A}$
that incorporates the possible differences in the measurement conditions.
Following \parref{Testing-if-the}, the likelihood-ratio log is $\Lambda=L\left(\boldsymbol{P}_{A},\boldsymbol{P}_{B}\right)-L\left(r,\boldsymbol{P}\right)$,
where 
\begin{equation}
L\left(\boldsymbol{P}_{A},\boldsymbol{P}_{B}\right)=\sum_{k=1}^{n_{A}}\ln\left(1+\boldsymbol{P}_{A}\cdot\bm{{\lambda}}_{k}\right)+\sum_{k=1}^{n_{B}}\ln\left(1+\boldsymbol{P}_{B}\cdot\bm{{\lambda}}_{k}\right),\label{eq:logLike-AB}
\end{equation}
and 
\begin{equation}
L\left(r,\boldsymbol{P}\right)=\sum_{k=1}^{n_{A}}\ln\left(1+\boldsymbol{P}\cdot\bm{{\lambda}}_{k}\right)+\sum_{k=1}^{n_{B}}\ln\left(1+r\boldsymbol{P}\cdot\bm{{\lambda}}_{k}\right).\label{eq:logLike-rB}
\end{equation}

$L\left(\boldsymbol{P}_{1,A},\boldsymbol{P}_{1,B}\right)$ can be
estimated easily by summing the log-likelihood for each data set at
its maximum, according to eq.~(\ref{eq:P1}). However, the approximation
of \subsecref{Analytical-approximation} for $L\left(r,\boldsymbol{P}\right)$,
yields a non-linear set of equations for the maximum:

\begin{eqnarray}
L\left(r,\boldsymbol{P}\right) & \simeq & \boldsymbol{P}\cdot\left(\boldsymbol{b}_{A}+r\boldsymbol{b}_{B}\right)-\left(\boldsymbol{P}\mathsf{J}_{A}\boldsymbol{P}+r^{2}\boldsymbol{P}\mathsf{J}_{B}\boldsymbol{P}\right)/2,\\
\partial_{r}L & \simeq & \boldsymbol{P}\cdot\boldsymbol{b}_{B}-r\boldsymbol{P}\mathsf{J}_{B}\boldsymbol{P}=0,\\
\nabla_{\boldsymbol{P}}L & \simeq & \boldsymbol{b}_{A}+r\boldsymbol{b}_{B}-\left(\mathsf{J}_{A}+r^{2}\mathsf{J}_{B}\right)\boldsymbol{P}=\boldsymbol{0}.
\end{eqnarray}
Searching for a 4-dimensional ($P_{x},P_{y},P_{z},r$) optimization
is CPU intensive. However, we can reduce the problem to a 1-dimensional
numerical optimization that requires a single loop over the sample,
by noting the log-likelihood ridge at 
\begin{equation}
\boldsymbol{P}_{0}\left(r\right)=\left(\mathsf{J}_{A}+r^{2}\mathsf{J}_{B}\right)^{-1}\left(\boldsymbol{b}_{A}+r\boldsymbol{b}_{B}\right),\label{eq:P0r}
\end{equation}
and recalling that $\boldsymbol{b}_{A}$, $\boldsymbol{b}_{B}$, $\mathsf{J}_{A}$
and $\mathsf{J}_{B}$ are available from the maximization of $L\left(\boldsymbol{P}_{A},\boldsymbol{P}_{B}\right)$,
leaving a dependence only on $r$. 

We maximize the likelihood by starting from $r_{0}=P_{1,B}/P_{1,A}$,
and taking Newton\textendash Raphson steps, \begin{linenomath}
\begin{equation}
r_{i+1}=r_{i}-\left[\nicefrac{\frac{\mathrm{d}L}{\mathrm{d}r}}{\frac{\mathrm{d}^{2}L}{\mathrm{d}r^{2}}}\right]_{\boldsymbol{P}=\boldsymbol{P}_{0}\left(r_{i}\right)}^{r=r_{i}},
\end{equation}
\end{linenomath} where\begin{linenomath}
\begin{equation}
\frac{\mathrm{d}L}{\mathrm{d}r}=\partial_{r}L+\nabla_{\boldsymbol{P}}L\cdot\partial_{r}\boldsymbol{P}\simeq\boldsymbol{P}\cdot\boldsymbol{b}_{B}-r\boldsymbol{P}\mathsf{J}_{B}\boldsymbol{P},\label{eq:drL4}
\end{equation}
\end{linenomath}and\begin{linenomath}
\begin{align}
\frac{\mathrm{d}^{2}L}{\mathrm{d}r^{2}} & \simeq\partial_{r}^{2}L+\nabla_{\boldsymbol{P}}\partial_{r}L\cdot\partial_{r}\boldsymbol{P}\nonumber \\
 & =-r\boldsymbol{P}\mathsf{J}_{B}\boldsymbol{P}+\left(\boldsymbol{b}_{B}-r\mathsf{J}_{B}\boldsymbol{P}\right)\cdot\left(\left(\mathsf{J}_{A}+r^{2}\mathsf{J}_{B}\right)^{-1}\boldsymbol{b}_{B}-2r\frac{\mathsf{J}_{A}^{-1}\mathsf{J}_{B}\mathsf{J}_{A}^{-1}\left(\boldsymbol{b}_{A}+r\boldsymbol{b}_{B}\right)}{\left(1+r^{2}\,\mathrm{trace}\left(\mathsf{J}_{A}^{-1}\mathsf{J}_{B}\right)\right)^{2}}\right).\label{eq:dr2L4}
\end{align}
\end{linenomath}As before, the approximations in eqs.~(\ref{eq:drL4}\textendash \ref{eq:dr2L4})
are those of \subsecref{Analytical-approximation}.

We define $\left(r_{1},\boldsymbol{P}_{1}\right)$ as the maximum
of the ridge as. The exact gradient of the log-likelihood is \begin{linenomath}
\begin{align}
\partial_{r}L & =\sum_{k=1}^{n_{B}}\frac{\boldsymbol{P}\cdot\bm{{\lambda}}_{k}}{1+r\boldsymbol{P}\cdot\bm{{\lambda}}_{k}},\nonumber \\
\nabla_{\boldsymbol{P}}L & =\sum_{k=1}^{n_{A}}\frac{\bm{{\lambda}}_{k}}{1+\boldsymbol{P}\cdot\bm{{\lambda}}_{k}}+r\sum_{k=1}^{n_{B}}\frac{\bm{{\lambda}}_{k}}{1+r\boldsymbol{P}\cdot\bm{{\lambda}}_{k}},
\end{align}
\end{linenomath}and the Hessian is\begin{linenomath}
\begin{equation}
H=-\sum_{k=1}^{n_{A}}\frac{\begin{pmatrix}0 & \boldsymbol{0}^{T}\\
\boldsymbol{0} & \bm{{\lambda}}_{k}\otimes\bm{{\lambda}}_{k}
\end{pmatrix}}{\left(1+\boldsymbol{P}\cdot\bm{{\lambda}}_{k}\right)^{2}}-\sum_{k=1}^{n_{B}}\frac{\begin{pmatrix}\left(\boldsymbol{P}\cdot\bm{{\lambda}}_{k}\right)^{2} & r\boldsymbol{P}\cdot\bm{{\lambda}}_{k}\otimes\bm{{\lambda}}_{k}\\
r\boldsymbol{P}\cdot\bm{{\lambda}}_{k}\otimes\bm{{\lambda}}_{k} & r^{2}\bm{{\lambda}}_{k}\otimes\bm{{\lambda}}_{k}
\end{pmatrix}}{\left(1+r\boldsymbol{P}\cdot\bm{{\lambda}}_{k}\right)^{2}}+\sum_{k=1}^{n_{B}}\frac{\begin{pmatrix}0 & \bm{{\lambda}}_{k}^{T}\\
\bm{{\lambda}}_{k} & 0_{3\times3}
\end{pmatrix}}{1+r\boldsymbol{P}\cdot\bm{{\lambda}}_{k}}.
\end{equation}
\end{linenomath}

Finally, the exact maximum is sought through $\binom{r_{i+1}}{\boldsymbol{P}_{i+1}}=\binom{r_{i}}{\boldsymbol{P}_{i}}-\left[H^{-1}\binom{\partial{}_{r}L}{\nabla_{\boldsymbol{P}}L}\right]_{\binom{r_{i}}{\boldsymbol{P}_{i}}}$,
and the \textquoteleft null hypothesis\textquoteright{} test is $p\left(2\Lambda,2\right)>5\%$.

This derivation can be extended for maximizing $L\left(r,\boldsymbol{P}\left(v\right)\right)$
and $L\left(r\left(v\right),\boldsymbol{P}\left(v\right)\right)$.

\subsection{Combining the method's tools\label{subsec:Combining-the-method's}}

In \subsecref{coordinate-system} we showed how to extract the components
of a constant polarization in spherical coordinates, independent of
any model. In \subsecref{formalism_for_varying_polarization} we showed
how to extract a varying polarization in Cartesian coordinates, again,
independent of any model. In \subsecref{Comparison-to-theoretical}
we showed how to compare a measurement with a model in Cartesian coordinates,
and proclaim that it could be extended to a varying description. 

In appendix~\ref{sec:Appendix-FullExample}, we demonstrate how these
three tools are combined to a continuous description of the ratio
between a measurement and its corresponding calculated prediction,
in spherical coordinates. Further, we provide in appendix~\ref{sec:Appendix-FullExample}
the full algorithm, which can be implemented in a computer program.

\section{Validity and advantages of the method\label{sec:Validity}}

In order to demonstrate the validity and advantages of this method,
we apply it to analyze simulated polarization-transfer data sets.
We produced two types of simulated events: 
\begin{enumerate}
\item Events with a constant polarization, which we analyzed according to
the procedures of \subsecref{Analytical-approximation} and \subsecref{coordinate-system}.
\item Events with polarization that depends on a kinematical parameter,
to demonstrate the continuous presentation discussed in \subsecref{formalism_for_varying_polarization}
and the modeling tools in \subsecref{Comparison-to-theoretical}.
\end{enumerate}
Thus, we test the usefulness of the MLEs, and demonstrate the advantages
of the maximization process and the presentation in spherical of coordinates.

\subsection{Simulation of a constant polarization\label{subsec:Validity-MC}}

In the first test we applied the new analysis method to simulated
data with a constant polarization to verify how well we reproduce
the \textquoteleft input\textquoteright{} polarization and to estimate
the inherent error of this method. 

\subsubsection{Generation of simulated events\label{subsec:Generation-of-simulated}}

We simulated 50,000 polarization-transfer experiments, where each
experiment consists of 100,000 events. The polarization was set to
be the same for all events in all the simulated experiments: $\boldsymbol{P}_{\mathrm{input}}=\left(-0.5,0,0.5\right)$.
The polarization of each event was multiplied by a constant analyzing
power of 0.25 in all the simulated experiments. The spin vector of
each event was rotated by a randomly generated spin precession matrix
($\mathsf{S}$) transforming the spin from target to a hypothetical
polarimeter (see \figref{Spin-precession-of}), where the three Euler
angles' distributions are \begin{linenomath}
\begin{align}
\alpha_{\mathrm{Euler}} & \sim\mathcal{N}\left(180^{\circ},15{}^{\circ}\right)\nonumber \\
\beta_{\mathrm{Euler}} & \sim\mathcal{N}\left(322^{\circ},15{}^{\circ}\right)\nonumber \\
\gamma_{\mathrm{Euler}} & \sim\mathcal{N}\left(90^{\circ},15{}^{\circ}\right).\label{eq:euler_dist}
\end{align}
\end{linenomath}Such distributions are typical to some experiments
using small solid angle spectrometers~\cite{Pospischil:2000pu,pospischil_thesis,Samo2017,Strauch,Doria2015,guy_ron_thesis,jlabDeep,ceepLet,deep2012PLB}.
The scattering by the carbon analyzer was simulated by assigning each
event an azimuthal angle $\phi_{\mathrm{FPP}}$ from a sinusoidal
distribution (eq.~(\ref{eq:dist})). 

For each one of the 50,000 simulated experiments, we extracted $\boldsymbol{P}_{\mathrm{exp}}$
and $\Delta\boldsymbol{P}_{\mathrm{exp}}$, by applying the new procedure
and analyzing the 100,000 transformed events (according to sections~\ref{subsec:Analytical-approximation}
and \ref{subsec:coordinate-system}). We define the error of each
experiment as $\boldsymbol{P}_{\mathrm{error}}\equiv\boldsymbol{P}_{\mathrm{exp}}-\boldsymbol{P}_{\mathrm{input}}$.
Notice that this error describes exactly the error of the procedure,
since neither the input events nor any of the simulated transformations
have any inherent errors and no errors in the measurements of the
angles or positions are assumed or applied (unlike a real simulation
of an experimental detector). 

To study our method's validity, we investigated the distributions
of $\boldsymbol{P}_{\mathrm{error}}$ and of the uncertainties $\Delta\boldsymbol{P}_{\mathrm{exp}}$
obtained for each data set.

\subsubsection{Results of the method applied to simulated events with constant polarization}

\begin{figure}
\begin{centering}
\includegraphics[width=0.9\columnwidth]{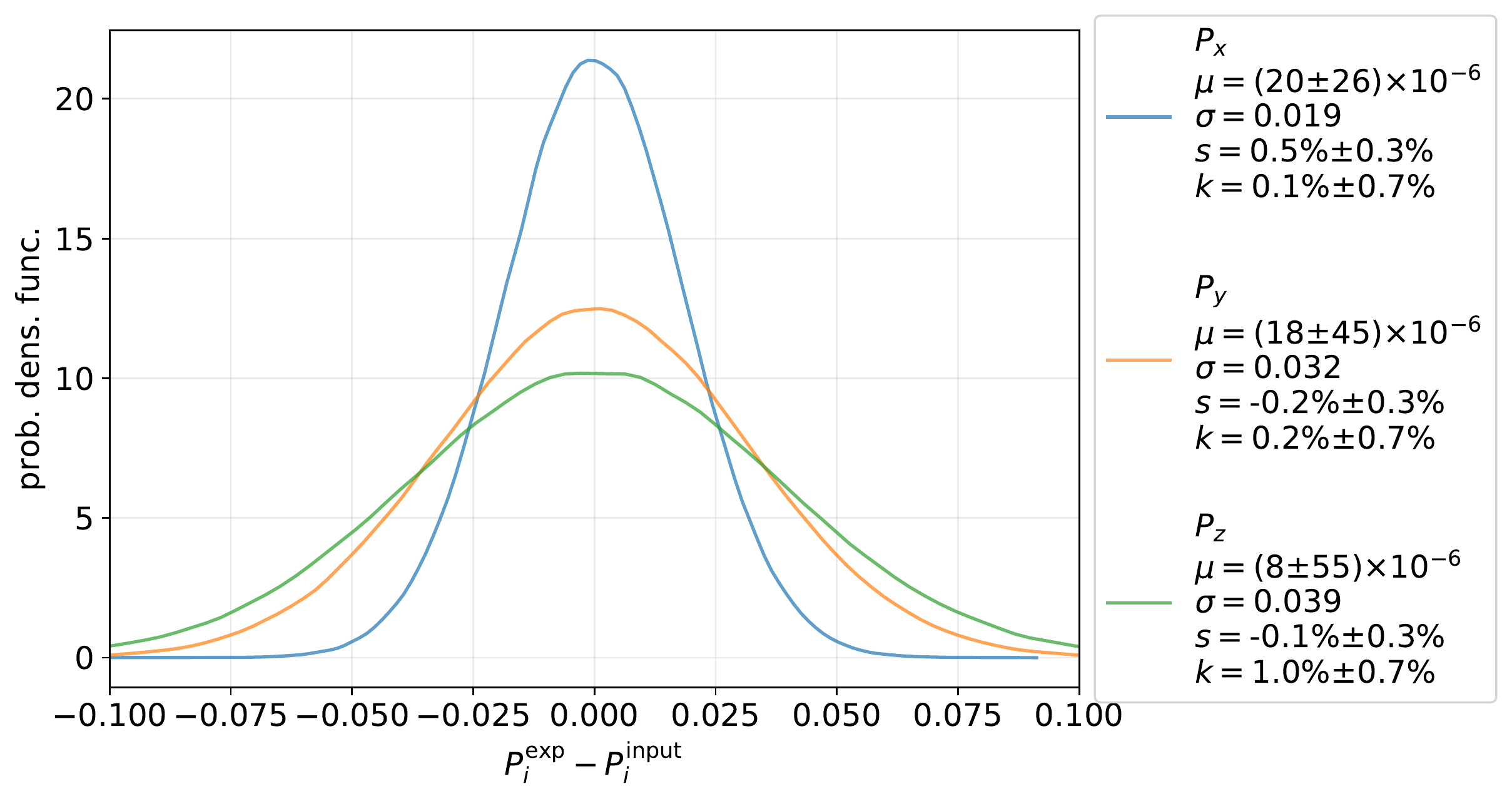}
\par\end{centering}
\caption{\label{fig:P_error}The distribution of the simulated polarization
component estimation errors: $P_{i}^{\mathrm{error}}\equiv P_{i}^{\mathrm{exp}}-P_{i}^{\mathrm{input}}$
for $i\in\left\{ x,y,z\right\} $. The legend shows the mean ($\mu$),
standard deviation ($\sigma$), skewness ($s$) and excess kurtosis
($k$) for each distribution.}
\end{figure}

The distributions of $P_{i}^{\mathrm{error}}$ ($i\in\left\{ x,y,z\right\} $)
for the 50,000 simulated experiments are shown in \figref{P_error}.
The mean value of each distribution $\left\langle P_{i}^{\mathrm{error}}\right\rangle $
is consistent with zero, as are the skewness and excess kurtosis of
each distribution. These results confirm that the MLEs for $\left(P_{x},P_{y},P_{z}\right)$
obtained by this analysis method are normally distributed and not
biased. 

\begin{figure}
\begin{centering}
\includegraphics[width=1\columnwidth]{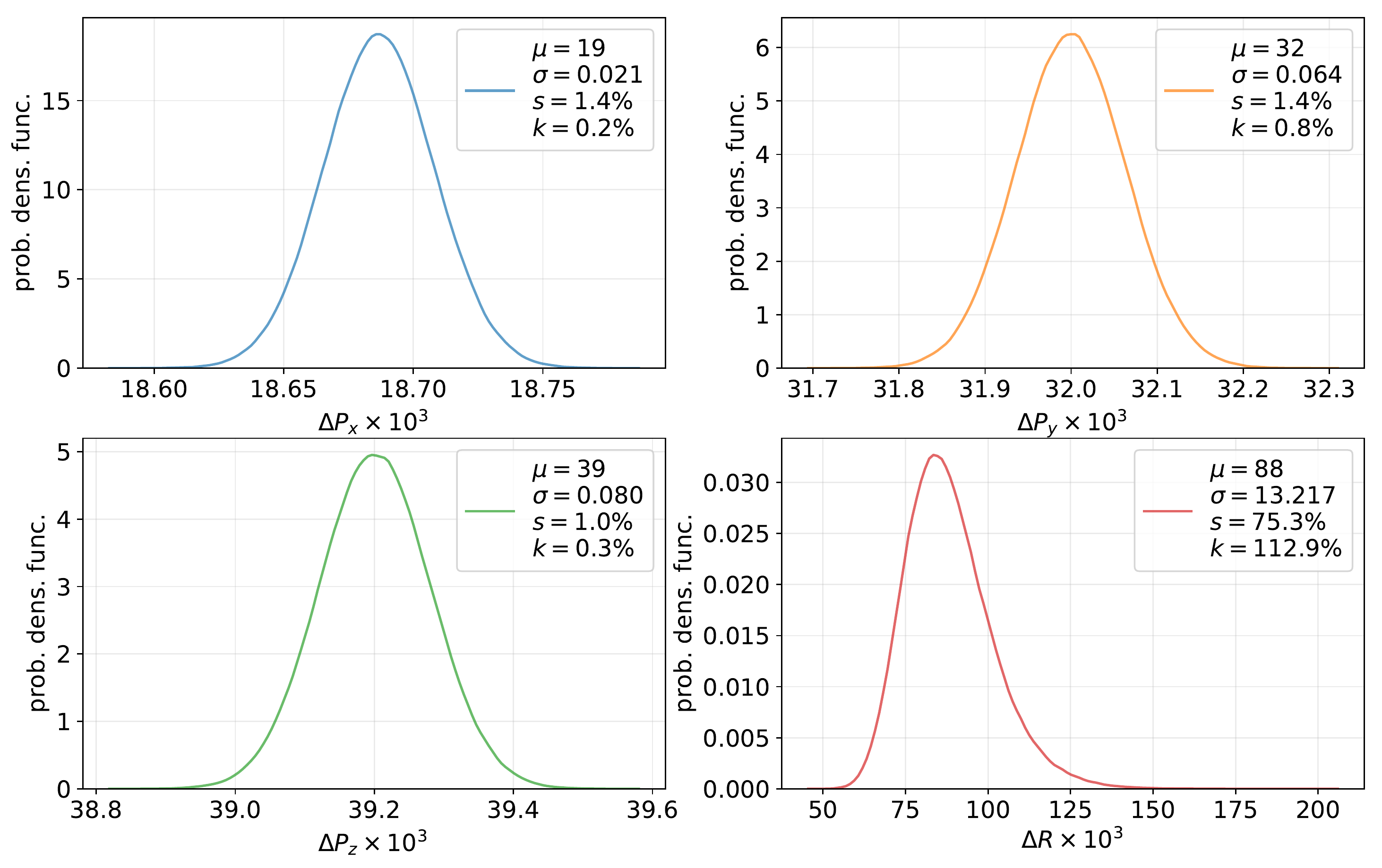}
\par\end{centering}
\caption{\label{fig:P_rel_unc_error}The distribution of the simulated polarization
component uncertainty estimators. The legend shows the mean ($\mu$),
standard deviation ($\sigma$), skewness ($s$) and excess kurtosis
($k$) for each distribution.}
\end{figure}

The distributions of the uncertainties $\Delta P_{i}^{\mathrm{exp}}$
($i\in\left\{ x,y,z\right\} $) obtained from each data set, are shown
in \figref{P_rel_unc_error}. The calculated means, $\left\langle \Delta P_{i}^{\mathrm{exp}}\right\rangle $,
are consistent with the standard deviations of the extracted polarization,
$\sigma\left[P_{i}^{\mathrm{exp}}\right]$. This consistency indicates
that the estimated uncertainties $\Delta P_{i}^{\mathrm{exp}}$ are
also unbiased in this method.

To assess the reliability of the relative uncertainties obtained from
the analysis, we compared the width of the error distribution to that
of the $P_{i}^{\mathrm{exp}}$ distribution by $\sigma\left[\Delta P_{i}^{\mathrm{exp}}\right]/\sigma\left[P_{i}^{\mathrm{exp}}\right]$.
The resultant estimated uncertainty on the calculated uncertainties
is \ensuremath{\sim} 0.2\% (see \figref{P_rel_unc_error}).

\begin{figure}
\begin{centering}
\includegraphics[width=0.9\columnwidth]{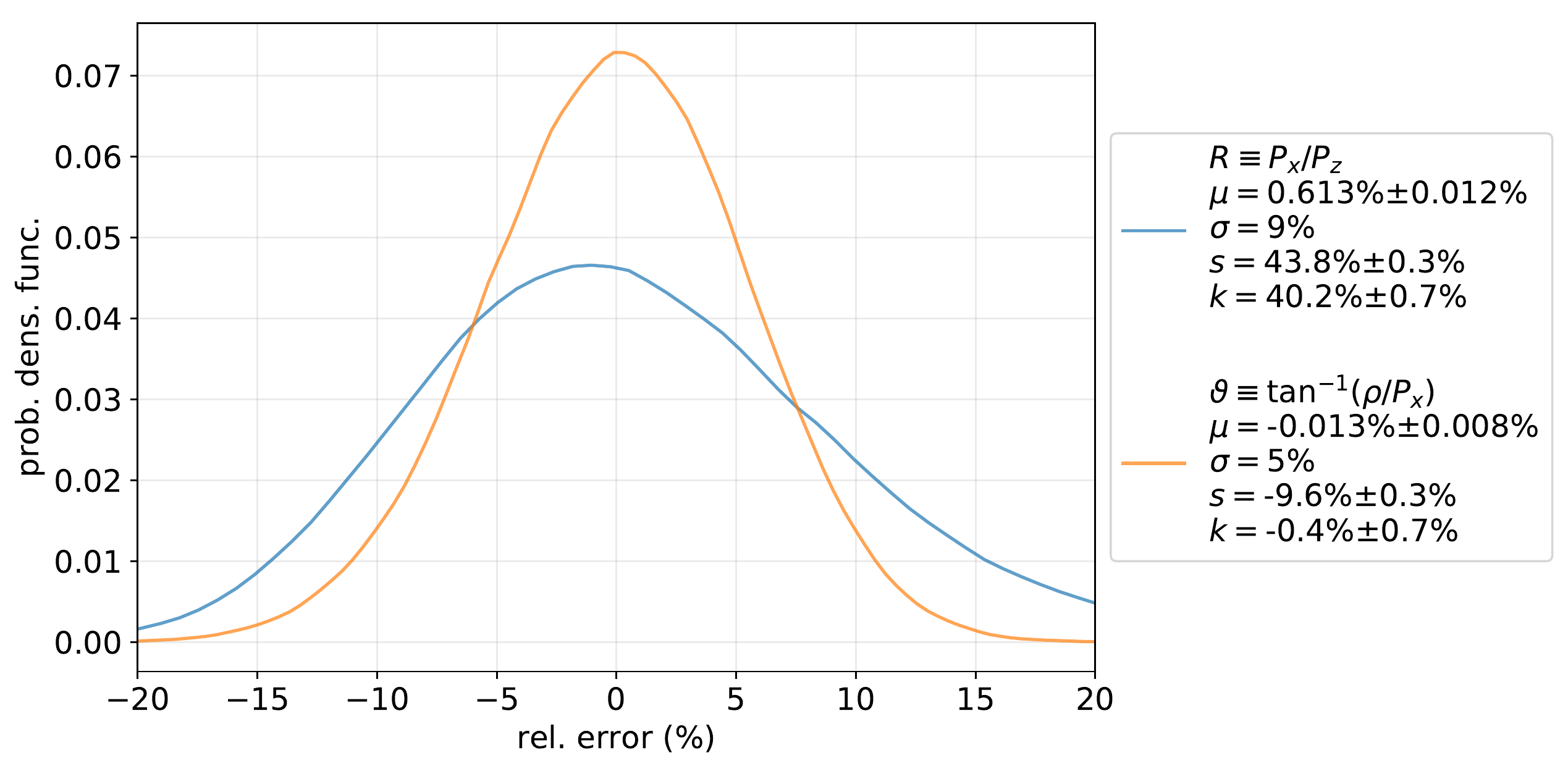}
\par\end{centering}
\caption{\label{fig:R_error}The distribution of the simulated polarization
estimation relative errors for $R$ and $\vartheta$: $\left(R_{\mathrm{exp}}-R_{\mathrm{input}}\right)/R_{\mathrm{input}}$
and $\left(\vartheta_{\mathrm{exp}}-\vartheta_{\mathrm{input}}\right)/\vartheta_{\mathrm{input}}$.
The legend shows the mean ($\mu$), standard deviation ($\sigma$),
skewness ($s$) and excess kurtosis ($k$) for each distribution.}
\end{figure}

While the determination of the individual components is obtained with
relatively small uncertainties as expected from the statistics, the
determination of component ratios results in large uncertainties,
which is typical to analyses using the standard method. This may be
an important issue in polarization experiments where beam or target
polarization are measured with relatively large uncertainties, which
in turn introduce large systematic errors on polarization components.
Determination of the component ratio such as $P_{x}/P_{z}$ largely
cancels this uncertainty and is used routinely to determine the electric
to magnetic form factors ratio. Thus we conclude with examining this
quantity in comparison to alternative ratios obtained when using spherical
coordinates.

The distribution of $R=\frac{P_{x}}{P_{z}}$ in the 50,000 experiments
shows that the estimator is biased ($\left\langle R_{\mathrm{exp}}-R_{\mathrm{input}}\right\rangle $
is inconsistent with zero), and has significantly large skewness and
excess kurtosis (see \figref{R_error}). This should be expected,
since for a bi-normal set $\left(x,y\right)$, $\mathrm{E}\left[\left|\frac{x}{y}\right|\right]>\frac{\mathrm{E}\left[\left|x\right|\right]}{\mathrm{E}\left[\left|y\right|\right]}$.
When extracting the propagated uncertainty of $R$ from the separate
polarization components, 
\[
\Delta R\equiv R\sqrt{\Delta P_{x}^{2}/P_{x}^{2}+\Delta P_{z}^{2}/P_{z}^{2}-2\,\mathrm{Cov}\left[P_{x},P_{z}\right]/P_{x}P_{z}}
\]
(eq.~(\ref{eq:dR})), it is found to be unbiased. However, the relative
uncertainty of the estimated uncertainty for $R$, estimated by evaluating
$\sigma\left[\Delta R\right]/\sigma\left[R\right]$, was found to
be $\sim15\%$, i.e.\ 100 times larger than those of the separate
components (see \figref{P_rel_unc_error}).

Unlike $R$, when using spherical coordinates (\subsecref{coordinate-system})
the MLE for $\vartheta$ (which comprises a similar physical meaning
to $R$, especially when $P_{y}=0$, see \subsecref{Motivation-spherical-coordinates})
is unbiased and almost normal (see \figref{R_error}). The relative\emph{
}uncertainty of the estimated uncertainty for $\vartheta$, estimated
by $\sigma\left[\Delta\vartheta\right]/\sigma\left[\vartheta\right]$,
is $\sim2\%$, much narrower than that of $R$. This better result
is due to avoiding propagation, as discussed in \subsecref{coordinate-system}.

In addition to generating narrower and more reliable uncertainties,
the simulation demonstrated the efficiency of our method regarding
CPU usage. In standard methods such as Minuit or Simplex, the typical
number of numerical steps is $\sim10$, and the likelihood function
is calculated between 60 and 100 times~\cite{deep2012PLB}. This
process can not be parallelized. In contrast, our method required
only a single Newton\textendash Raphson step to achieve convergence
in each of the 50,000 experiments. Therefore, our method requires
approximately 100 times less CPU time with respect to standard methods.

\subsection{Simulation of events with varying polarization}

We turn to demonstrate the advantage of the continuous presentation
of polarization that depends on a certain parameter. For this purpose
we produced events with a polarization that depends on a kinematical
variable, and analyzed them according to the procedures of sections~\ref{subsec:formalism_for_varying_polarization}
and~\ref{subsec:Comparison-to-theoretical}. Thus, we survey the
usefulness of a cubic spline description.

\subsubsection{Simulation and results}

For simplicity, we analyze a simulated data set whose polarization
depends on a single parameter. This was done by assuming that in the
simulated experiment a proton's polarization is known but the measured
polarization is scaled by the \textquoteleft system\textquoteright{}
analyzing power which depends on the proton momentum. Thus the measurement
should yield the momentum dependence of the analyzing power.

We simulated one polarization measurement experiment of 300,000 events.
The polarization was set to be the same for all events: $\boldsymbol{P}_{\mathrm{input}}=\left(-0.4,0.1,0.6\right)$.
For each event the proton was assigned a kinetic energy ($T_{p}$)
generated from a normal distribution. The polarization of each event
was multiplied by an analyzing power ($a_{y}$) that depends on $T_{p}$.
As in the previous section, the spin vector of each event was rotated
by a randomly generated spin precession matrix ($\mathsf{S}$) transforming
the spin from the target to the polarimeter, where the three Euler
angles' distributions are given in eq.~(\ref{eq:euler_dist}). The
scattering of each event by the polarimeter analyzer was simulated
by assigning each event an azimuthal angle $\phi_{\mathrm{FPP}}$
from a sinusoidal distribution (eq.~(\ref{eq:dist})). Further details
can be found in the \href{http://www-nuclear.tau.ac.il/~ecohen/PublishedPapers/polar2/Polar2.html}{simulation documentation}.
\begin{figure}[b]
\begin{centering}
\includegraphics[width=0.8\columnwidth]{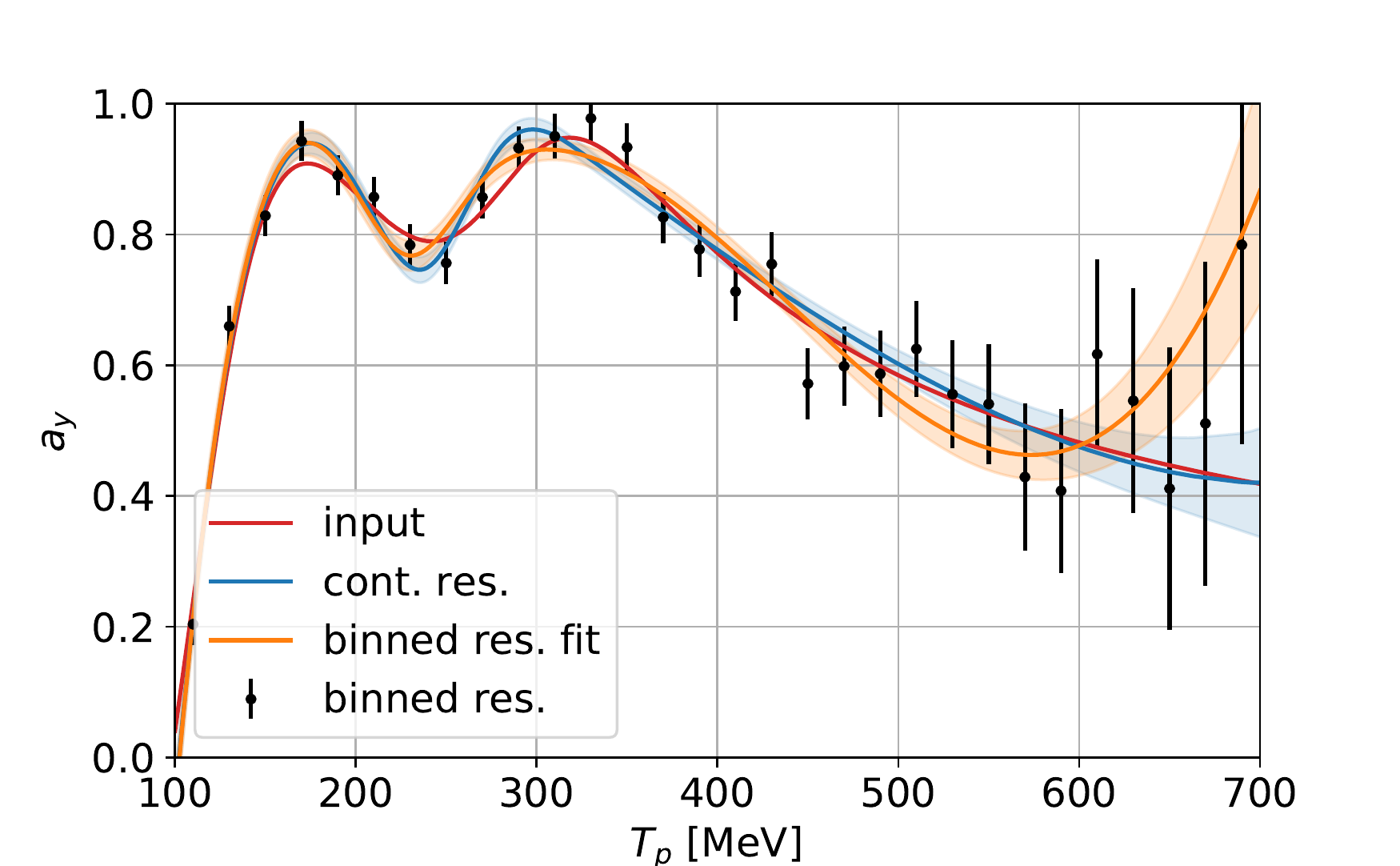}
\par\end{centering}
\caption{\label{fig:Simulated-analyzing-power.}Simulated analyzing power.
We compare the input form (red line), with binned (black error-bars)
and continuous (blue band) analyses. For a full comparison, we fitted
the binned results to a cubic spline with the same number of parameters
as the continuous description (orange band). See the \protect\href{http://www-nuclear.tau.ac.il/~ecohen/PublishedPapers/polar2/Polar2.html}{simulation documentation}
for additional details and fits.}
\end{figure}

We analyzed the data in the ``traditional way'', in bins of kinetic
energy (\subsecref{Binned-likelihood}). The analysis yields the ratio
of the measured polarization components to those of the constant polarization
considered event by event and averaged over each bin. It represents
the average analyzing power over the bin. The analysis was repeated
using a cubic spline representation of the $T_{p}$ dependence of
the measured $a_{y}$, applying the MLE approach according to \subsecref{Cubic-spline-interpolation}.

In \figref{Simulated-analyzing-power.} we compare the results of
both methods. We note that the spline result shows the full structure
of the input function, and does not exhibit any false structure. This
cannot be stated about the binned results where one may be tempted
to associate statistical fluctuations with some extra structure, as
suggested by the fit in \figref{Simulated-analyzing-power.}. One
may be concerned that addition of more parameters to the spline will
cause false structure to emerge due to over-fitting. However, using
Wilks' theorem for the likelihood-ratio, one can show that extra parameters
are redundant as discussed in \subsecref{Comparison-to-theoretical}
and the \href{http://www-nuclear.tau.ac.il/~ecohen/PublishedPapers/polar2/Polar2.html}{simulation documentation}.
We also note that the continuous presentation results in an uncertainty
band which is much narrower than the uncertainties in the binned analysis
(where each bin is independent of its neighbors) particularly in the
bins with the low statistics. 

We can use the same simulation to show that if in the above problem
the analyzing power is known up to a unknown factor, we can reduce
the uncertainty of the estimated factor by $\sim15\%$ by using the
method of \subsecref{Quantification-of-the}.

\subsubsection{Documented source code notebook}

The detailed simulation description accompanied with source code snippets
is \href{http://www-nuclear.tau.ac.il/~ecohen/PublishedPapers/polar2/Polar2.html}{available on-line}. 

The source code ``Jupyter'' notebook can be downloaded from \href{http://www-nuclear.tau.ac.il/~ecohen/PublishedPapers/polar2/Polar2.ipynb}{here}
(file size = 300 kB, run time $\approx$ 5 seconds).

The notebook contains the implementation of the majority of the formul\ae\ 
from \secref{The-method}.

\section{Summary and Conclusions\label{sec:Summary-and-Conclusions}}

We introduced a new method to analyze polarization measurement data.
The analysis is performed using spherical coordinates and a continuous
presentation of the measurements. The spherical coordinates ($P$,
$\rho$ and $\vartheta$) are unaffected by mixing of the $P_{z}$
and $P_{y}$ components which in turn affects the uncertainty in the
measurements of the components ratio. Using $\vartheta$ instead of
$R$ does not change the physical interpretation. However, if one
needs an observable that is only a function of $R$, we can redefine
eq.~(\ref{eq:S}) so that $\tan\vartheta\equiv P_{x}/P_{z}$ without
changing the formalism. 

The continuous presentation (rather than dividing the measured range
into arbitrary bins in which discrete average values are determined,
independent of each other) results in an uncertainty band which is
narrower than the errors in a binned analysis, particularly when some
bins are measured with lower statistics. It also allows a reliable
comparison of the measurement to other measurements or calculations.
The new method yields even smaller uncertainties which in the simulated
examples were reduced by about 20\% (see \figref{Simulated-analyzing-power.}).
When the data are described by well known models or parameterizations
(like a dipole form factor) the uncertainty can be reduced even further
(see \subsecref{Comparison-to-theoretical}). 

Our new method is more efficient in CPU by about two orders of magnitude,
when compared with conventional methods currently in use.

\acknowledgments

The authors would like to thank Dr.~Linda Montag for her help in
the manuscript preparation.

This work is supported by the Israel Science Foundation (Grant 390/15)
of the Israel Academy of Arts and Sciences, by the Israel Atomic Energy
Commission (the PAZI Foundation), and by the Israel Ministry of Science,
Technology and Space. \newpage{}

\appendix

\section*{Appendix}

\section{Full Example\label{sec:Appendix-FullExample}}

As discussed in \subsecref{Combining-the-method's}, in \subsecref{coordinate-system}
we use spherical coordinates for the simplest case. In \subsecref{formalism_for_varying_polarization}
we deal with varying polarization only in Cartesian coordinates, and
in \subsecref{Comparison-to-theoretical} we compare a measurement
with a model only in Cartesian coordinates. 

Here we demonstrate how these three tools are combined to a continuous
description of the ratio between a measurement and its corresponding
calculated prediction, in spherical coordinates. Further, we provide
the full algorithm, which can be implemented in a computer program.

\subsection{Derivation}

We measure the polarization-transfer in the $^{2}\mathrm{H}\left(\vec{e},e'\vec{p}\right)n$
reaction at a fixed beam energy, as a function of the missing momentum
$p_{m}$. We would like to account for the finite acceptance in $Q^{2}$,
$\theta_{pq}$ and $\phi_{pq}$, while comparing with a model that
predicts that $P_{y}=0$ (and therefore $\varphi_{\mathrm{mod}}=0$).

We divide the $p_{m}$ range to $N$ segments with $3\left(N+1\right)$
parameters according to \subsecref{Cubic-spline-interpolation}.

Here we take the short hand notation of $w_{l}=w_{l}\left(p_{m}\right)$,
and define $r\left(p_{m}\right)\equiv P_{\mathrm{exp}}\left(p_{m}\right)/P_{\mathrm{calc}}\left(p_{m}\right)$,
$\delta\left(p_{m}\right)\equiv\vartheta_{\mathrm{exp}}\left(p_{m}\right)-\vartheta_{\mathrm{\mathrm{calc}}}\left(p_{m}\right)$.
Since $\varphi_{\mathrm{\mathrm{calc}}}=0$, we will use $\varphi\equiv\varphi_{\mathrm{exp}}$.

$\gamma$ is:\begin{linenomath}
\begin{eqnarray}
\gamma & = & P_{\mathrm{exp}}\left(\lambda_{x}\cos\vartheta_{\mathrm{exp}}+\sin\vartheta_{\mathrm{exp}}\left(\lambda_{y}\sin\varphi+\lambda_{z}\cos\varphi\right)\right)\nonumber \\
 & = & r\,P_{\mathrm{\mathrm{calc}}}\left(\lambda_{x}\cos\left(\vartheta_{\mathrm{\mathrm{calc}}}+\delta\right)+\sin\left(\vartheta_{\mathrm{mod}}+\delta\right)\left(\lambda_{y}\sin\varphi+\lambda_{z}\cos\varphi\right)\right)\nonumber \\
 & = & \left(\sum_{l=0}^{N}w_{l}r_{l}\right)P_{\mathrm{\mathrm{calc}}}\times\nonumber \\
 &  & \times\left(\lambda_{x}\cos\left(\vartheta_{\mathrm{\mathrm{calc}}}+\sum_{l=0}^{N}w_{l}\delta_{l}\right)+\sin\left(\vartheta_{\mathrm{\mathrm{calc}}}+\sum_{l=0}^{N}w_{l}\delta_{l}\right)\times\right.\nonumber \\
 &  & \;\;\;\;\;\;\;\;\;\left.\times\left(\lambda_{y}\sin\sum_{l=0}^{N}w_{l}\varphi_{l}+\lambda_{z}\cos\sum_{l=0}^{N}w_{l}\varphi_{l}\right)\right)
\end{eqnarray}
\end{linenomath}The the first derivatives of $\gamma$ are:\begin{linenomath}
\begin{align}
\partial_{r_{l}}\gamma & =\frac{w_{l}\gamma}{\sum_{l'}w_{l'}r_{l'}}\equiv w_{l}\lambda'_{r}\nonumber \\
\partial_{\delta_{l}}\gamma & =w_{l}\left(\tau\xi-\lambda_{x}\rho\right)\equiv w_{l}\lambda'_{\delta}\nonumber \\
\partial_{\varphi_{l}}\gamma & =w_{l}\bar{\xi}\equiv w_{l}\lambda'_{\varphi},
\end{align}
\end{linenomath}where $\tau$, $\xi$, $\bar{\xi}$ and $\rho$ are
all functions of $p_{m}$ and are model dependent (as specified in
eq.~(\ref{eq:PxPyPz}) below). The second derivatives of $\gamma$
are:\begin{linenomath}
\begin{align}
\partial_{r_{l}}\partial_{r_{l'}}\gamma & =0\nonumber \\
\partial_{\delta_{l}}\partial_{\delta_{l'}}\gamma & =-w_{l}w_{l'}\gamma\equiv-w_{l}w_{l'}\lambda'_{\delta\delta}\nonumber \\
\partial_{\varphi_{l}}\partial_{\varphi_{l'}}\gamma & =-w_{l}w_{l'}\xi\equiv-w_{l}w_{l'}\lambda'_{\varphi\varphi}\nonumber \\
\partial_{r_{l}}\partial_{\delta_{l'}}\gamma & =\frac{w_{l}w_{l'}}{\sum_{l"}w_{l"}r_{l"}}\left(\tau\xi-\lambda_{x}\rho\right)\equiv-w_{l}w_{l'}\lambda'_{r\delta}\nonumber \\
\partial_{r_{l}}\partial_{\varphi_{l'}}\gamma & =\frac{w_{l}w_{l'}}{\sum_{l"}w_{l"}r_{l"}}\bar{\xi}\equiv-w_{l}w_{l'}\lambda'_{r\varphi}\nonumber \\
\partial_{\delta_{l}}\partial_{\varphi_{l'}}\gamma & =w_{l}w_{l'}\tau\bar{\xi}\equiv-w_{l}w_{l'}\lambda'_{\delta\varphi}.
\end{align}
\end{linenomath}

\subsection{Algorithm}
\begin{enumerate}
\item We divide the events to $N$ bins, and calculate the cubic spline
weights according to eq.~(\ref{eq:cubic_spline_ws}). 
\item We start from $r_{l}=1\,,\,\delta_{l}=\varphi_{l}=0$.
\item For each event we \label{enu:For-each-event}
\begin{enumerate}
\item calculate \begin{linenomath}
\begin{align}
\sum_{w}r & \equiv\sum_{l=0}^{N}w_{l}r_{l},\nonumber \\
\rho & =P_{\mathrm{\mathrm{calc}}}\sin\left(\vartheta_{\mathrm{\mathrm{calc}}}+\sum_{l=0}^{N}w_{l}\delta_{l}\right)\sum_{w}r,\nonumber \\
\xi & =\rho\left(\lambda_{y}\sin\sum_{l=0}^{N}w_{l}\varphi_{l}+\lambda_{z}\cos\sum_{l=0}^{N}w_{l}\varphi_{l}\right),\nonumber \\
\bar{\xi} & =\rho\left(\lambda_{y}\cos\sum_{l=0}^{N}w_{l}\varphi_{l}-\lambda_{z}\sin\sum_{l=0}^{N}w_{l}\varphi_{l}\right),\nonumber \\
\tau & =\cot\left(\vartheta_{\mathrm{\mathrm{calc}}}+\sum_{l=0}^{N}w_{l}\delta_{l}\right),\nonumber \\
\gamma & =\xi+\lambda_{x}P_{\mathrm{\mathrm{calc}}}\cos\left(\vartheta_{\mathrm{\mathrm{calc}}}+\sum_{l=0}^{N}w_{l}\delta_{l}\right)\sum_{w}r,\label{eq:PxPyPz}
\end{align}
\end{linenomath}
\item followed by 
\begin{equation}
\begin{pmatrix}\lambda'_{rk}\\
\lambda'_{\delta k}\\
\lambda'_{\varphi k}
\end{pmatrix}=\begin{pmatrix}\gamma/\sum_{w}r\\
\tau\xi-\lambda_{x}\rho\\
\bar{\xi}
\end{pmatrix},
\end{equation}
and 
\begin{equation}
\begin{pmatrix}\lambda'_{r\delta k}\\
\lambda'_{\delta\delta k}\\
\lambda'_{\delta\varphi k}\\
\lambda'_{r\varphi k}\\
\lambda'_{\varphi\varphi k}
\end{pmatrix}=\begin{pmatrix}\left(\lambda_{x}\rho-\tau\xi\right)/\sum_{w}r\\
\gamma\\
-\tau\bar{\xi}\\
-\bar{\xi}/\sum_{w}r\\
\xi
\end{pmatrix},
\end{equation}
\item which are then multiplied for each $l$ for $\frac{w_{lk}}{1+\gamma_{k}}\begin{pmatrix}\lambda'_{rk}\\
\lambda'_{\delta k}\\
\lambda'_{\varphi k}
\end{pmatrix}$, 
\item which is outer-squared for $\frac{w_{lk}w_{l'k}}{\left(1+\gamma_{k}\right)^{2}}\begin{pmatrix}\lambda_{rk}'^{2} & \lambda'_{rk}\lambda'_{\delta k} & \lambda'_{rk}\lambda'_{\varphi k}\\
\lambda'_{rk}\lambda'_{\delta k} & \lambda_{\delta k}'^{2} & \lambda'_{\delta k}\lambda'_{\varphi k}\\
\lambda'_{rk}\lambda'_{\varphi k} & \lambda'_{\delta k}\lambda'_{\varphi k} & \lambda_{\varphi k}'^{2}
\end{pmatrix}$, 
\item and finally for each $ll'$ pair we produce $\frac{w_{lk}w_{l'k}}{1+\gamma_{k}}\begin{pmatrix}0 & \lambda'_{r\delta k} & \lambda'_{r\varphi k}\\
\lambda'_{r\delta k} & \lambda'_{\delta\delta k} & \lambda'_{\delta\varphi k}\\
\lambda'_{r\varphi k} & \lambda'_{\delta\varphi k} & \lambda'_{\varphi\varphi k}
\end{pmatrix}$.
\end{enumerate}
\item We take a Newton\textendash Raphson step by \label{enu:We-take-a-Newton-Raphson-step}
by:
\begin{enumerate}
\item summing for the sub-vectors $\bar{\boldsymbol{b}}''_{l}$ and sub-matrices
$\bar{\mathrm{I}}''_{ll'}$ according to
\begin{equation}
\bar{\boldsymbol{b}}''_{l}=\sum_{k=1}^{n}\frac{w_{lk}}{1+\gamma_{k}}\begin{pmatrix}\lambda'_{rk}\\
\lambda'_{\delta k}\\
\lambda'_{\varphi k}
\end{pmatrix},
\end{equation}
and
\begin{equation}
\bar{\mathsf{J}}''_{ll'}=\sum_{k=1}^{n}\frac{w_{lk}w_{l'k}}{\left(1+\gamma_{k}\right)^{2}}\begin{pmatrix}\lambda_{rk}'^{2} & \lambda'_{rk}\lambda'_{\delta k} & \lambda'_{rk}\lambda'_{\varphi k}\\
\lambda'_{rk}\lambda'_{\delta k} & \lambda_{\delta k}'^{2} & \lambda'_{\delta k}\lambda'_{\varphi k}\\
\lambda'_{rk}\lambda'_{\varphi k} & \lambda'_{\delta k}\lambda'_{\varphi k} & \lambda_{\varphi k}'^{2}
\end{pmatrix}+\sum_{k=1}^{n}\frac{w_{lk}w_{l'k}}{1+\gamma_{k}}\begin{pmatrix}0 & \lambda'_{r\delta k} & \lambda'_{r\varphi k}\\
\lambda'_{r\delta k} & \lambda'_{\delta\delta k} & \lambda'_{\delta\varphi k}\\
\lambda'_{r\varphi k} & \lambda'_{\delta\varphi k} & \lambda'_{\varphi\varphi k}
\end{pmatrix},
\end{equation}
\item concatenating $\bar{\boldsymbol{b}}''_{l}$ and $\bar{\mathsf{J}}''_{ll'}$
for all $l,l'\in\left[0,N\right]$ to obtain the full $\boldsymbol{b}''$
and $\mathsf{J}''$,
\item and advancing $\boldsymbol{r}_{i}=\boldsymbol{r}_{i-1}+\mathrm{\mathsf{J}''}^{-1}\boldsymbol{b}''$,
where $\boldsymbol{r}\equiv\left(r_{0},\delta_{0},\varphi_{0},...,r_{N},\delta_{N},\varphi_{N}\right)^{T}$.
\end{enumerate}
\item We repeat the two previous steps (steps \ref{enu:For-each-event}
and \ref{enu:We-take-a-Newton-Raphson-step}) until $\left|\boldsymbol{r}_{i}-\boldsymbol{r}_{i-1}\right|<\varepsilon$
\& $\left|L_{i}-L_{i-1}\right|<\varepsilon$, where $\varepsilon$
is the ``machine's accuracy'', and $L=\sum_{k=1}^{n}\ln\left(1+\gamma_{k}\right)$.
\end{enumerate}
\newpage{}

\bibliographystyle{elsarticle-num}
\phantomsection\addcontentsline{toc}{section}{\refname}\bibliography{polar2}
\clearpage{}

\newpage{}\newpage{}\newpage{}

\includepdf[pages=-,width=0.9\paperwidth,height=1\paperheight,keepaspectratio]{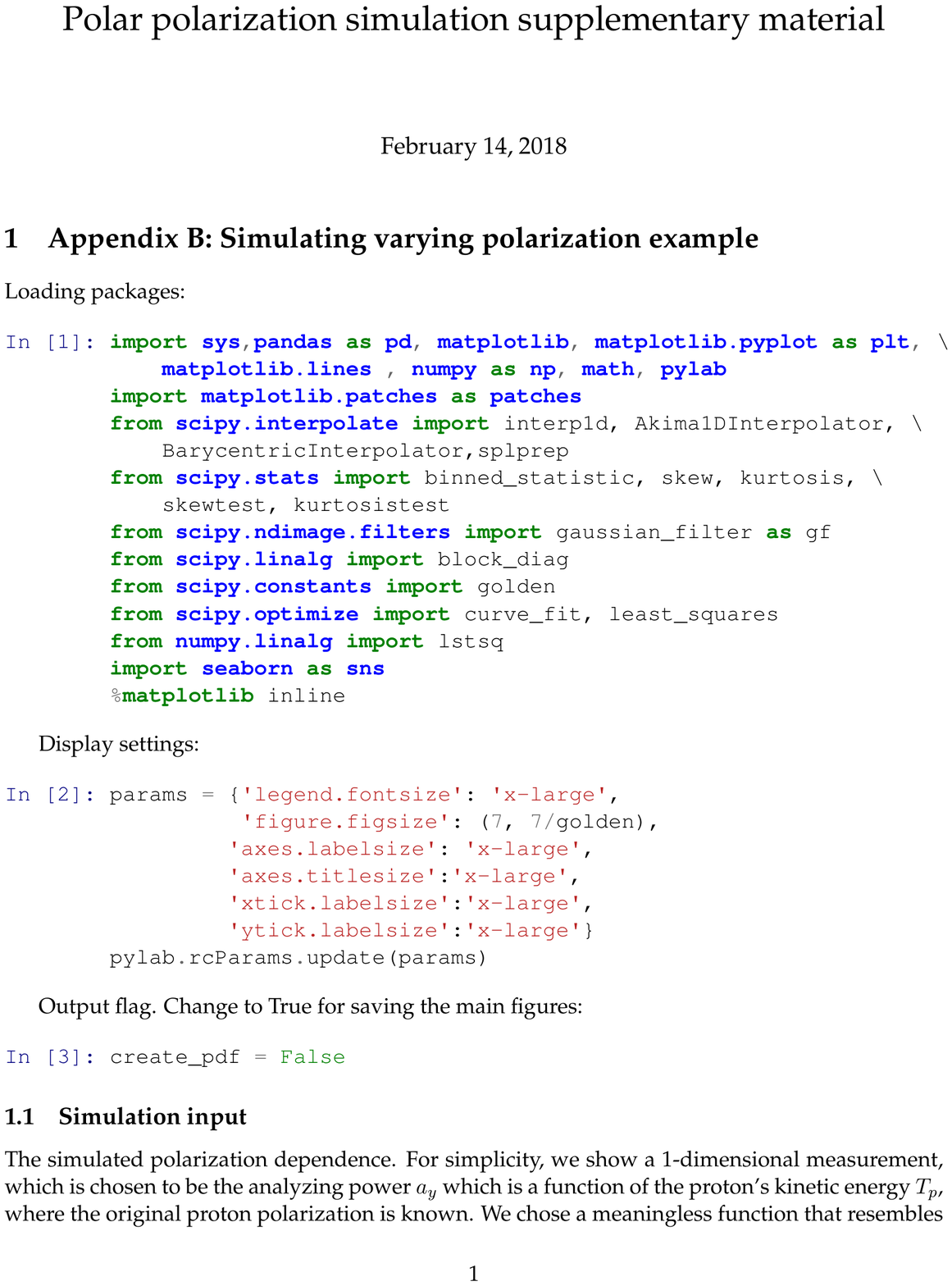}
\end{document}